\documentclass[review]{elsarticle}

\usepackage{lineno,hyperref}
\usepackage{physics}
\usepackage{amsmath,amssymb}
\usepackage{graphicx}
\usepackage{epstopdf, epsfig}
\usepackage{subcaption}

\modulolinenumbers[5]

\journal{arXiv}

\begin{document}

\begin{frontmatter}

\title{Rotating electroosmotic flow of power-law fluid through polyelectrolyte grafted microchannel}

\author{Maneesh Patel}
\author{S S Harish Kruthiventi}
\author{P Kaushik \corref{mycorrespondingauthor}}
\address{Department of Mechanical Engineering,\\ National Institute of Technology Tiruchirappalli,\\ Tamil Nadu 620015, India}


\cortext[mycorrespondingauthor]{Corresponding author}
\ead{pkaushik@nitt.edu}


\begin{abstract}
Due to evergrowing importance of understanding flow of bio-fluids in Lab-on-CD based systems, we investigate the flow behaviour of power-law fluids in the rotating electroosmotic flow through a polyelectrolyte grafted (soft) narrow channel. We use a in-house numerical code to solve the governing transport equations for the velocities and flow rates in a rotating channel subjected to an applied external electric field. We show the strong effect of polyelectrolyte layer on the flow behaviour and find an increase in flow rate as we increase the size of the polyelectrolyte layer. We also show that rheology strongly influences the interplay of the Coriolis forces due to rotation and electrical body force due to the applied electric field. We show that the velocities are generally higher for shear thinning fluids as compared to shear thickening fluids. We also show that presence of polymer brushes in the polyelectrolyte layer creates a drag on the fluid which reduces velocities.  We elaborate that the flow rates are strongly altered by the effect of rotation and that shear thickening fluids have lower flow rates than shear thinning fluids. We believe that studying effect of fluid rheology becomes very important for designing soft channel based Lab-on-CD systems driven by electroosmotic forcing and dealing with rheologically complex bio-fluids such as blood, saliva or mucus. 

\end{abstract}


\end{frontmatter}


\section{Introduction}

In the modern day, owing to the vast number of applications of micro-electromechanic systems (MEMS) such as lab-on-CD (LOCD) based devices, a lot of interest has been seen among resesarch community studying fluid flow in these devices \cite{integratedmfdevices}. The flow of fluid occrus in such LOCD based devices due to the rotation of the platform on which the channel may be etched. Flow augmentation and alteration may occur due to the presence of centrifugal and coriolis forces exerted by the rotation of the device. Further augmentation of such flows is often done by using electroosmotic forcing of the fluid. When a polar fluid is placed on a charged surface, there is ion redistribution within the fluid. Redistribution of ions causes a change in the charge distribution. Electroosmotic flow (EOF) is induced by applying an electric field across the channel to drive the ions of the polar fluid, which drag the fluid along with it. The theory of EOF has been well studied in literature \cite{stone2004engineering}. More recently, alteration to EOF has been done by grafting a layer of polyelectrolyte (PE) brushes to the walls of the microchannel. These brushes change the distribution of free ions within the bulk of the fluid and thereby changing the charge distribution \cite{chandasinhadas}. The layer containing PE brushes is assumed to allow flow of fluid through them, however, an extra drag term is added to flow governing equations to account for the physical obstruction to the flow of fluid by the PE brushes. The flow of fluids through such PE grafted micro and nano channels have been recently studied in a plethora of literature. Harden et al. \cite{Hardenetal2001} were one of the earliest researchers to report on the EOF of fluids through PE grafted microchannels giving insight into the mobility of fluid flow in the PE brush region. Controlling of electroosmotic flow by coating of PE layer was also reported \cite{Cao2011, Monteferrante2015}. We also find in the literature, reports on control of PE brushes using electric field and consequently the flow of fluid in PE grafted channels \cite{cao2012}. More recently variational methods have been used to solve the fluid flow equations in PE grafted microchannels \cite{Sadeghi2018}. Sterric and charge redistribution effects on flow through PE grafted channel was studied by Reshadi and Saidi \cite{RESHADI2018443}. In a recent study, the effect of fluid rheology was studied on the EOF through PE grafted soft channel by Gaikwad et al. \cite{gaikwad2018softness}; the authors found that the fluid rheology significantly impacts the net throughput of the fluid through PE grafted nanochannels. \\
The fundamental idea to enhance fluid flow using electroosmotic forcing in rotating microchannel may be attributed to Chang and Wang \cite{changandwang}; the authors considered fluid to be Newtonian in behaviour and found the existence of secondary flow velocities due to coriolis effect. Effect of transient and startup flow behaviour of rotational EOF has been studied analytically \cite{si_jian_chang_liu_2016, Gheshlaghi2016}. Rheological effects on the rotational EOF have been reported recently for power law fluid \cite{XIE2014231}, third grade fluid \cite{LI2015240}, viscoelastic fluid \cite{kaushikmandalchakraborty2017, ABHIMANYU201656}, Eyring fluid \cite{Qicheng2017} and viscoplastic material \cite{QI2017355}; the papers show that rheology of the fluid significantly alters the flow behaviour of fluid in rotating EOF. In particular, Abhimanyu et al. \cite{ABHIMANYU201656} show that EOF and rotational coriolis effects cannot be linearly superimposed when both are studied separately for rheologically complex fluids. In order to study the effect of secondary velocity, confining the flow laterally becomes important. The effect of lateral confinement was discussed in detail for Newtonian fluid by Ng and Qi \cite{ng2015electro} , power law fluid by Kaushik et al. \cite{Kaushik2017} and viscoelastic fluid by Kaushik et al. \cite{KAUSHIK2017123}; the authors found recirculation loops using streamlines which showed the importance of rotational flow as a method of inducing mixing withing the fluid. In more recent studies the effect of PE layer grafting on rotational EOF of Newtonian fluid for different charge distributions were studied by Kaushik et al. \cite{kaushikp2019} and Liu and Jian \cite{Liu2019}. Kaushik et al. \cite{kaushikp2019} found a significant enhancment in the flow velocities with increasing the size of the PE grafted layer in line with the net throughput increase due to the presence of PE grafted layer as reported by Gaikwad et al. \cite{gaikwad2018softness}. \\
Due to demand for modern bio-medical applications to be coupled with LOCD based devices, the requirement to understand the flow of bio-fluids in LOCD devices becomes important. Bio-fluids such as blood, salive and mucus are rheologically complex in nature and have to be modeled and understood accordingly. Although some work has been reported on flow of rheologically complex fluids through PE grafted micro and nano channels, no study has comprehensively reported the behaviour of rheologically complex fluids on the rotational EOF through PE grafted soft microchannels. Accordingly, in the present study, we attempt to solve the governing equations numerically for flow of a power-law fluid in a rotating channel with flow enhanced by electroosmotic forcing. We try to validate the results from the present study with data available in literature for the special case of Newtonian fluid as well as channel without PE grafting. Further, we study the effect of various parameters such as graft layer size, the drag coefficient, rotational velocity and power-law index on the flow behaviour.

\section{Mathematical Formulation}

\begin{figure}
  \centerline{\includegraphics[width=100mm]{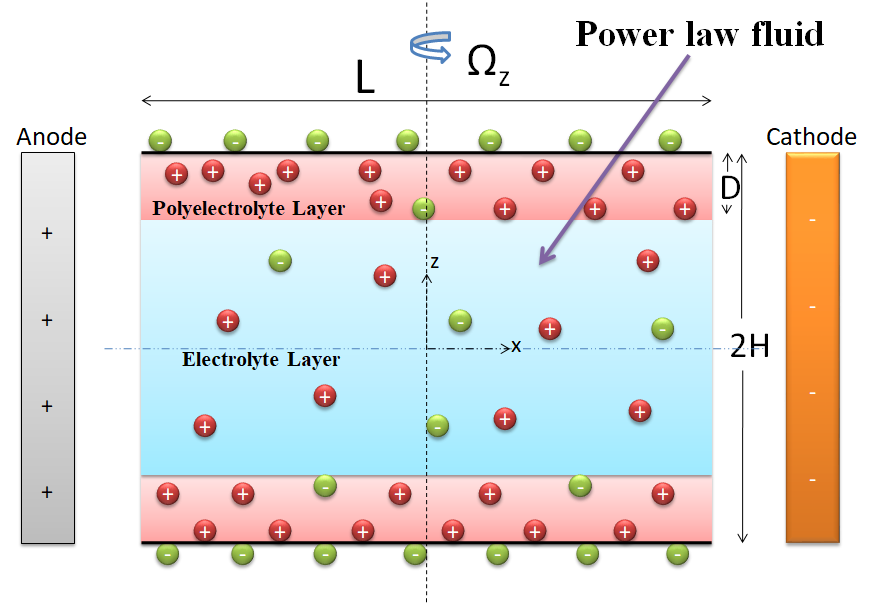}}
\setlength{\abovecaptionskip}{0pt plus 0pt minus 0pt}
  \caption{A schematic representation of the fluid domain and configuration.}
\label{fig:k1}
\end{figure}

As shown in figure \ref{fig:k1}, let us consider a power-law fluid is confined between two parallel plates of length $L$ and width $W$ at a distance $2H$ apart. An electric field of magnitude $E$ is applied in the $x$-direction. The plates are rotated with an angular velocity $\Omega_z$ in the $z$-direction. The thickness of the PE grafted layer is given by $D$. The coordinate system is chosen in such a way that the origin lies at the centre of the channel as seen from figure \ref{fig:k1}. \\

The governing equations of fluid flow are
\begin{equation} \label{cont_eqn}
\div \vec V = 0
\end{equation}
and
\begin{equation} \label{el_mom_eqn}
\rho \left[ \frac{\partial \vec V}{\partial t} + \left( \vec V \cdot \boldsymbol{\nabla} \right) \vec V + 2 \vec \Omega \cross \vec V \right] = -\boldsymbol{\nabla} P + \div \boldsymbol{\tau} + \vec F, 0 \le z \le H-D
\end{equation}
\begin{equation} \label{pel_mom_eqn}
\rho \left[ \frac{\partial \vec V}{\partial t} + \left( \vec V \cdot \boldsymbol{\nabla} \right) \vec V + 2 \vec \Omega \cross \vec V \right] = -\boldsymbol{\nabla} P + \div \boldsymbol{\tau} - \mu_c V_i{^n} + \vec F, H-D \le z \le H
\end{equation}

The Cauchy momentum equation above has been written separately for the electrolyte layer (EL) given by equation \ref{el_mom_eqn} and the PE grafted layer (PEL) given by equation \ref{pel_mom_eqn} to incorporate the extra drag term on the fluid velocities in the PEL given by $\mu_c \vec V$. In equations \ref{el_mom_eqn} and \ref{pel_mom_eqn}, $\rho$ is the density of the fluid, $\vec V$ is the velocity, symbol $t$ represents time, $\vec \Omega$ is the angular velocity of the channel given by $\vec \Omega = \left[ 0, 0, \Omega_z \right]$, $\mu_c$ is the drag coeffient in the PEL, $P$ is the modified pressure incorporating the centrifugal term given by $P=p-\rho \left|\vec \Omega \cross \vec r \right|^2/2$, $\boldsymbol{\tau}$ is the stress tensor and $\vec F$ is the body force term which will be used in our further analysis to incorporate the electroosmotic body force term. We simplify the stress tensor for power-law fluid following Bird et al. \cite{transphen} to get,
\begin{equation} \label{pl_stress}
\boldsymbol{\tau} = \eta \left( \left| \dot \gamma \right| \right)  \left[ \boldsymbol{\nabla} \vec V + \left( \boldsymbol{\nabla} \vec V \right)^T  \right]
\end{equation}
where, $\left| \dot \gamma \right|$ is the magnitude rate of strain tensor given $\left| \dot \gamma \right|= \left[ \dfrac{1}{2} \dot \gamma : \dot \gamma  \right]^{\dfrac{1}{2}}$, the strain tensor $\dot \gamma = \left[ \boldsymbol{\nabla} \vec V + \left( \boldsymbol{\nabla} \vec V \right)^T  \right]$  and $\eta \left( \dot \gamma  \right) = k_{pl} \left| \dot \gamma \right| ^{n-1}$, with $k_{pl}$ being known as the flow consistency index and $n$ is known as the power index of the power-law fluid. \\

It is important to mention here that it is the coriolis force that causes the flow in transverse direction and the centrifugal force only acts as aiding the existing pressure gradient \cite{changandwang}. In the present work, following Kaushik et al. \cite{kaushikp2019}, we assume no externally applied pressure gradient to drive the flow; we also assume that $L \gg H$, using which we drop the pressure gradient term in our subsequent analysis. We would also like to mention that the drag coefficient within the PEL, given by $\mu_c$ is proportional to ${\left( n-1 \right)}^{th}$ power of the velocity component following Parnas and Cohen \cite{parnas1987}.

In order to incorporate the electrical body force term, we assume that the PE grafted channel is filled with a liquid of dielectric constant $\epsilon$ with $\pm z_v$ being the valence of cations and anions, respectively within the liquid. The number density of cations and anions withing the liquid is given by $m_\pm$. The thickness of the PE brush layer is assumed to be fixed and the number density of ions are assumed to not vary with the electrostatic effects and pH of the liquid. The existence of PE brushes is also assumed to not alter the permittivity within PE grafted layer. An external electric field $E_x$ is applied along the $x$-direction. To solve for the flow velocites it is essential to solve for the charge distribution within the two layers in which fluid flow is considered. Following Gaikwad et al. \cite{gaikwad2018softness}, the governing equations for charge distribution $\left( \psi \right)$ within the two layers of the channel are given by
\begin{equation} \label{el_charge_eqn}
\dfrac{d^2 \psi}{d z^2}= \dfrac{-e z_v \left( m_+ - m_- \right)}{\epsilon}, 0 \le z \le H-D
\end{equation}
\begin{equation} \label{pel_charge_eqn}
\dfrac{d^2 \psi}{d z^2}= \dfrac{-e z_v \left( m_+ - m_- \right)+Z_v e M}{\epsilon}, H-D \le z \le H
\end{equation}
where, $e$ represents the charge of an electron, $Z_v$ represents the valence of ions within the PE grafted layer and $M$ is the number density of ions within the PE grafted layer.\\

On simplification of the governing equations \ref{cont_eqn}, \ref{el_mom_eqn} and \ref{pel_mom_eqn} using the earlier mentioned assumptions, we get the $x$-momentum and $y$ momentum equations in EL (valid in the region $0 \le z \le H-D$) as:
\begin{equation} \label{el_xmom_eqn}
\rho \left[ \dfrac{\partial u}{\partial t} - 2 \Omega_z v \right] = \dfrac{\partial}{\partial z} \left( \eta \left( \left| \dot \gamma \right| \right) \dfrac{\partial u}{\partial z} \right) + e z_v \left( m_+-m_- \right) E_x
\end{equation}
\begin{equation} \label{el_ymom_eqn}
\rho \left[ \dfrac{\partial v}{\partial t} + 2 \Omega_z u \right] = \dfrac{\partial}{\partial z} \left( \eta \left( \left| \dot \gamma \right| \right) \dfrac{\partial v}{\partial z} \right)
\end{equation}
for PEL (valid in the region $H-D \le z \le H$), we get,
 \begin{equation} \label{pel_xmom_eqn}
\rho \left[ \dfrac{\partial u}{\partial t} - 2 \Omega_z v \right] = \dfrac{\partial}{\partial z} \left( \eta \left( \left| \dot \gamma \right| \right) \dfrac{\partial u}{\partial z} \right) - \mu_c u^n + e z_v \left( m_+-m_- \right) E_x
\end{equation}
\begin{equation} \label{pel_ymom_eqn}
\rho \left[ \dfrac{\partial v}{\partial t} + 2 \Omega_z u \right] = \dfrac{\partial}{\partial z} \left( \eta \left( \left| \dot \gamma \right| \right) \dfrac{\partial v}{\partial z} \right) - \mu_c v^n
\end{equation}

The boundary conditions for the charge distribtion and momentum equations in the EL and PEL are as follows: \\
At the symmetry line $ \left( z=0 \right)$, $\dfrac{\partial \psi}{\partial z} =0$, $\dfrac{\partial u}{\partial z} =0$ and $\dfrac{\partial v}{\partial z} =0$\\
At the wall $ \left( z=H \right)$, $\dfrac{\partial \psi}{\partial z} =0$ (Gaussian boundary condition) and $u=v=0$ (no-slip condition) \\
At the interace between EL and PEL $ \left( z=H-D \right)$, $\psi_{EL}=\psi_{PEL}$, $\left[ \dfrac{\partial \psi}{\partial z}\right]_{EL} =\left[\dfrac{\partial \psi}{\partial z}\right]_{PEL}$, $u_{EL} = u_{PEL}$, $\left[ \dfrac{\partial u}{\partial z}\right]_{EL} =\left[\dfrac{\partial u}{\partial z}\right]_{PEL}$, $v_{EL}=v_{PEL}$ and $\left[ \dfrac{\partial v}{\partial z}\right]_{EL} =\left[\dfrac{\partial v}{\partial z}\right]_{PEL}$ \\

We assume that the fluid is at rest initially, however, charge is distributed as per the governing equations \ref{el_charge_eqn} and \ref{pel_charge_eqn}. \\

The charge distribution given by equations \ref{el_charge_eqn} and \ref{pel_charge_eqn} is assumed to follow the Boltzmann distribution, given by, $m_\pm = m_\infty \exp{\mp \dfrac {e z_v \psi}{k_B T}}$, $k_B$ being the Boltzmann constant and $T$ being the temperature. Using the Debye-Huckel linearization, we may write $\exp{\mp \dfrac {e z_v \psi}{k_B T}} \approx \left( 1\mp \dfrac {e z_v \psi}{k_B T} \right) $ to get the charge distribution equations in the EL and PEL respectively as
\begin{equation} \label{el_lambda_charge_eqn}
\dfrac{d^2 \psi}{d z^2}= \dfrac{\psi}{\lambda ^2},  0 \le z \le H-D
\end{equation}
\begin{equation} \label{pel_lambda_charge_eqn}
\dfrac{d^2 \psi}{d z^2}= \dfrac{\psi}{\lambda ^2}-\dfrac{\psi_s}{\lambda_p ^2},  H-D \le z \le H
\end{equation}
In the above equations \ref{el_lambda_charge_eqn} and \ref{pel_lambda_charge_eqn}, the electric double layer (EDL) thickness within the EL is given by $\lambda =\sqrt {\dfrac{\epsilon k_B T}{2 m_\infty e^2 z_v^2}}$ and EDL thickness within the PEL is given by $\lambda_p = =\sqrt {\dfrac{\epsilon k_B T}{2 M e^2 z_v Z_v}}$ and $\psi_s = \dfrac{e Z_v}{k_B T}$. \\
Equations \ref{el_lambda_charge_eqn} and \ref{pel_lambda_charge_eqn} are nondimensionalized by using $\psi^* = \dfrac{\psi}{\psi_{ref}}$ where $\psi_{ref}=\dfrac{e z_v}{k_B T}$, dimensionless length scale $z^* = \dfrac{z}{H}$ giving $d=\dfrac{D}{H}$, the dimensionless inverse of the EDL thickness in the EL $\kappa = \dfrac{H}{\lambda}$ and in the PEL $\kappa_p = \dfrac{H}{\lambda_p}\sqrt{\dfrac{\psi_s}{\psi_{ref}}}$. This yields
\begin{equation} \label{el_dless_charge_eqn}
\dfrac{d^2 \psi^*}{d z^*{^2}}= \kappa^2 \psi^*,  0 \le z \le 1-d
\end{equation}
\begin{equation} \label{pel_dless_charge_eqn}
\dfrac{d^2 \psi^*}{d z^*{^2}}= \kappa^2 \psi^*-\kappa_p^2,  1-d \le z \le 1
\end{equation}
In order to non-dimensionalize the momentum equations \ref{el_xmom_eqn}, \ref{el_ymom_eqn}, \ref{pel_xmom_eqn} and \ref{pel_xmom_eqn}, we use, the velocity scale as the Smoluchowski velocity of a power-law fluid \cite{ZHAO2008503, gaikwad2018softness}, given by $u_s = n \left( \dfrac{1}{\lambda} \right)^{\dfrac{1-n}{n}} \left(-\dfrac{\epsilon \psi_{ref} E_x}{k_{pl}}   \right)^{\dfrac{1}{n}}$ and shear stress scale as $\tau_{ref} = -\dfrac{\epsilon \psi_{ref} E_x}{\lambda}$ to get the dimensionless $x$-momentum and $y$ momentum equations in EL (valid in the region $0 \le z \le 1-d$) as:
\begin{equation} \label{el_dless_xmom_eqn}
\left[ \dfrac{\partial u^*}{\partial t^*} - 2 Re_\Omega v^* \right] = \dfrac{\partial}{\partial z^*} \left( \eta^* \left( \left| \dot \gamma^* \right| \right) \dfrac{\partial u^*}{\partial z^*} \right) + \kappa^2 \psi^*
\end{equation}
\begin{equation} \label{el_dless_ymom_eqn}
\left[ \dfrac{\partial v^*}{\partial t^*} + 2 Re_\Omega u^* \right] = \dfrac{\partial}{\partial z^*} \left( \eta^* \left( \left| \dot \gamma^* \right| \right) \dfrac{\partial v^*}{\partial z^*} \right)
\end{equation}
for PEL (valid in the region $1-d \le z \le 1$), we get,
 \begin{equation} \label{pel_dless_xmom_eqn}
\left[ \dfrac{\partial u^*}{\partial t^*} - 2 Re_\Omega v^* \right] = \dfrac{\partial}{\partial z^*} \left( \eta^* \left( \left| \dot \gamma^* \right| \right) \dfrac{\partial u^*}{\partial z^*} \right) - \alpha^2 u^*{^n} + \kappa^2 \psi^*
\end{equation}
\begin{equation} \label{pel_dless_ymom_eqn}
\left[ \dfrac{\partial v^*}{\partial t^*} + 2 Re_\Omega u^* \right] = \dfrac{\partial}{\partial z^*} \left( \eta^* \left( \left| \dot \gamma^* \right| \right) \dfrac{\partial v^*}{\partial z^*} \right) - \alpha^2 v^*{^n}
\end{equation}
In the above equations \ref{el_dless_xmom_eqn}, \ref{el_dless_ymom_eqn}, \ref{pel_dless_xmom_eqn} and \ref{pel_dless_ymom_eqn}, $u^* = \dfrac{u}{u_s}$, $v^* = \dfrac{v}{u_s}$, $t^* = \dfrac{t \epsilon \psi_{ref} E_x}{\rho \lambda u_s H}$, $Re_\Omega = \dfrac{\rho \Omega_z u_s H^2}{\epsilon \psi_{ref} E_x}$, $\eta^* = n^n {\dot \gamma \lambda}^{n-1}$, $\dot \gamma^* = \dot \gamma \left( \dfrac{H}{u_s} \right)$ and $\alpha = H \sqrt{\dfrac{n^n \mu_c \lambda}{k_{pl}}}$. 
The boundary conditions for the dimensionless charge distribtion and momentum equations in the EL and PEL are as follows: \\
At the symmetry line $ \left( z^*=0 \right)$, $\dfrac{\partial \psi^*}{\partial z^*} =0$, $\dfrac{\partial u^*}{\partial z^*} =0$ and $\dfrac{\partial v^*}{\partial z^*} =0$\\
At the wall $ \left( z^*=1 \right)$, $\dfrac{\partial \psi^*}{\partial z^*} =0$ (Gaussian boundary condition) and $u^*=v^*=0$ (no-slip condition) \\
At the interace between EL and PEL $ \left( z^*=1-d \right)$, $\psi^*_{EL}=\psi^*_{PEL}$, $\left[ \dfrac{\partial \psi^*}{\partial z^*}\right]_{EL} =\left[\dfrac{\partial \psi^*}{\partial z^*}\right]_{PEL}$, $u^*_{EL} = u^*_{PEL}$, $\left[ \dfrac{\partial u^*}{\partial z^*}\right]_{EL} =\left[\dfrac{\partial u^*}{\partial z^*}\right]_{PEL}$, $v^*_{EL}=v^*_{PEL}$ and $\left[ \dfrac{\partial v^*}{\partial z^*}\right]_{EL} =\left[\dfrac{\partial v^*}{\partial z^*}\right]_{PEL}$ \\

To ease subsequent discussion on to solution method and results, we drop the superscript $*$ from the dimensionless equations \ref{el_dless_charge_eqn}, \ref{pel_dless_charge_eqn}, \ref{el_dless_xmom_eqn}, \ref{el_dless_ymom_eqn}, \ref{pel_dless_xmom_eqn} and \ref{pel_dless_ymom_eqn}. \textit{The terms without superscript $*$ henceforth will mean dimensionless quantities}.

\section{Solution method and validation}

For the present problem, first the dimensionless charge distribution equations \ref{el_dless_charge_eqn} and \ref{pel_dless_charge_eqn} are solved analytically following Kaushik et al. \cite{kaushikp2019} with boundary conditions to get,
\begin{equation} \label{psi_el_soln}
{ \psi}=C\cosh{kz} , 0 \le z \le 1-d
\end{equation}
\begin{equation} \label{psi_pel_soln}
{\psi}=A\cosh{kz}+B\sinh{kz}+\beta^2, 1-d \le z \le 1
\end{equation}
where, $\beta = \dfrac{\kappa_p}{\kappa}$, $B=\dfrac{\beta^2 \tanh{kL}}{\cosh{kL} \left[ 1-\tanh^2{kL} \right] }$, $A=-B\dfrac{\cosh{k}}{\sinh{k}}$ \\ and $C=-B\left[ {\dfrac{cosh{k}}{sinh{k}}-\dfrac{\cosh{kL}}{\sinh{kL}}} \right]$. \\

In order to solve the Cauchy momentum equations \ref{el_dless_xmom_eqn}, \ref{el_dless_ymom_eqn}, \ref{pel_dless_xmom_eqn} and \ref{pel_dless_ymom_eqn}, an in-house finite difference code is developed. We use the forward in time finite difference and central difference for space approach to descretize the governing equations. Marching forward in time is done by fully implicit method. The solution is considered converged when steady state is reached (i.e. no signifcant change in the velocity profiles). The values of the velocities at a point $z_i$ is given by $u_i$ and $v_i$. The momentum equations for the power law fluid can be approximated by following difference equations:

\begin{equation}\label{fd_el_u_eqn}
\dfrac{u^{q+1}_i-u^{q}_i}{\Delta t}-2 Re_\Omega v^{q+1}_i=\dfrac{ \eta^{q+1}_{i+\frac{1}{2}}(u^{q+1}_{i+1}-u^{q+1}_i)- \eta^{q+1}_{i-\frac{1}{2}}(u^{q+1}_{i}-u^{q+1}_{i-1})}{\Delta z^2}+k^2\psi^2_i
\end{equation}

\begin{equation}\label{fd_el_v_eqn}
\dfrac{v^{q+1}_i-v^{q}_i}{\Delta t}+2 Re_\Omega u^{q+1}_i=\dfrac{ \eta^{q+1}_{i+\frac{1}{2}}(v^{q+1}_{i+1}-v^{q+1}_i)- \eta^{q+1}_{i-\frac{1}{2}}(v^{q+1}_{i}-v^{q+1}_{i-1})}{\Delta z^2}
\end{equation}

\begin{equation}\label{fd_pel_u_eqn}
\dfrac{u^{q+1}_i-u^{q}_i}{\Delta t}-2 Re_\Omega v^{q+1}_i=\dfrac{ \eta^{q+1}_{i+\frac{1}{2}}(u^{q+1}_{i+1}-u^{q+1}_i)- \eta^{q+1}_{i-\frac{1}{2}}(u^{q+1}_{i}-u^{q+1}_{i-1})}{\Delta z^2}+k^2\psi^2_i-\alpha^2 {(u^{n-1})}^{q}_i u^{q+1}_i
\end{equation}

\begin{equation}\label{fd_pel_v_eqn}
\dfrac{v^{q+1}_i-v^{q}_i}{\Delta t}+2 Re_\Omega u^{q+1}_i=\dfrac{ \eta^{q+1}_{i+\frac{1}{2}}(v^{q+1}_{i+1}-v^{q+1}_i)- \eta^{q+1}_{i-\frac{1}{2}}(v^{q+1}_{i}-v^{q+1}_{i-1})}{\Delta z^2}-\alpha^2{(v^{n-1})}^{q}_i v^{q+1}_i
\end{equation}

Where $q$ denotes the time level and $ \eta_{i\pm \frac{1}{2}}= \left( \eta_{i\pm1}+ \eta_i \right) /2$ denotes the shear strain rate used in momentum equation and is given by:
\begin{equation} \label {dless_strain1}
|\dot\gamma_i|=\sqrt{\left( \frac{u_{i+1}-u_{i-1}}{2\Delta z}\right)^2+\left(\frac{u_{i+1}-u_{i-1}}{2\Delta z}\right)^2}
\end{equation}

To solve the momentum equation by numerical method we use initial guess value as the velocities obtained from the previous time step. For the given guess value, we get the strain rate from equation \ref {dless_strain1}. The effective viscosity is calculated and then used to obtain matrix form of the system of difference equations. This is further solved. \\

\begin{figure}
\begin{center}
        \begin{subfigure}[t]{0.5\textwidth}
                \includegraphics[width=1\linewidth]{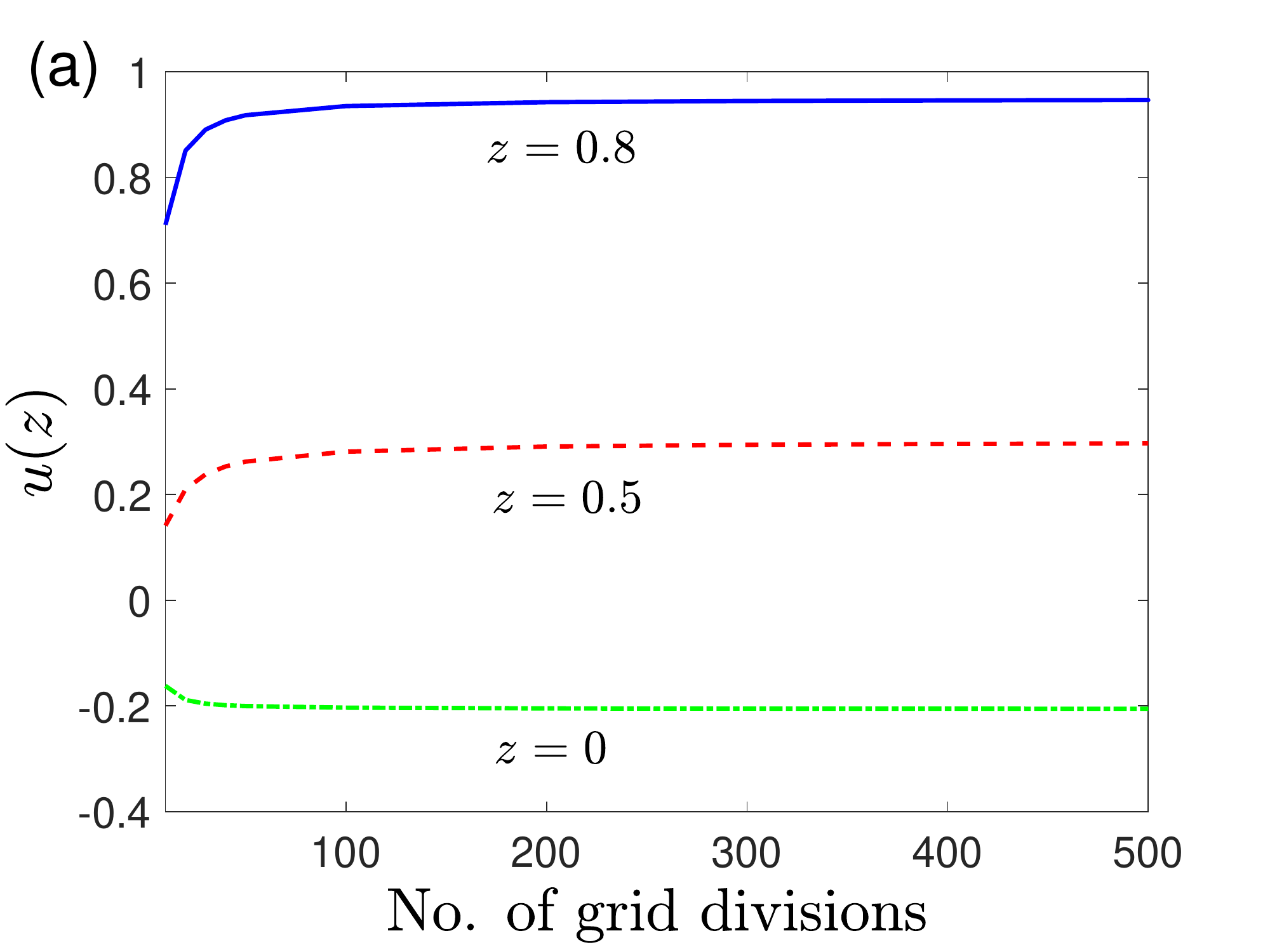}
                \label{fig:k2a}
        \end{subfigure}\hfill
        \begin{subfigure}[t]{0.5\textwidth}
                \includegraphics[width=1\linewidth]{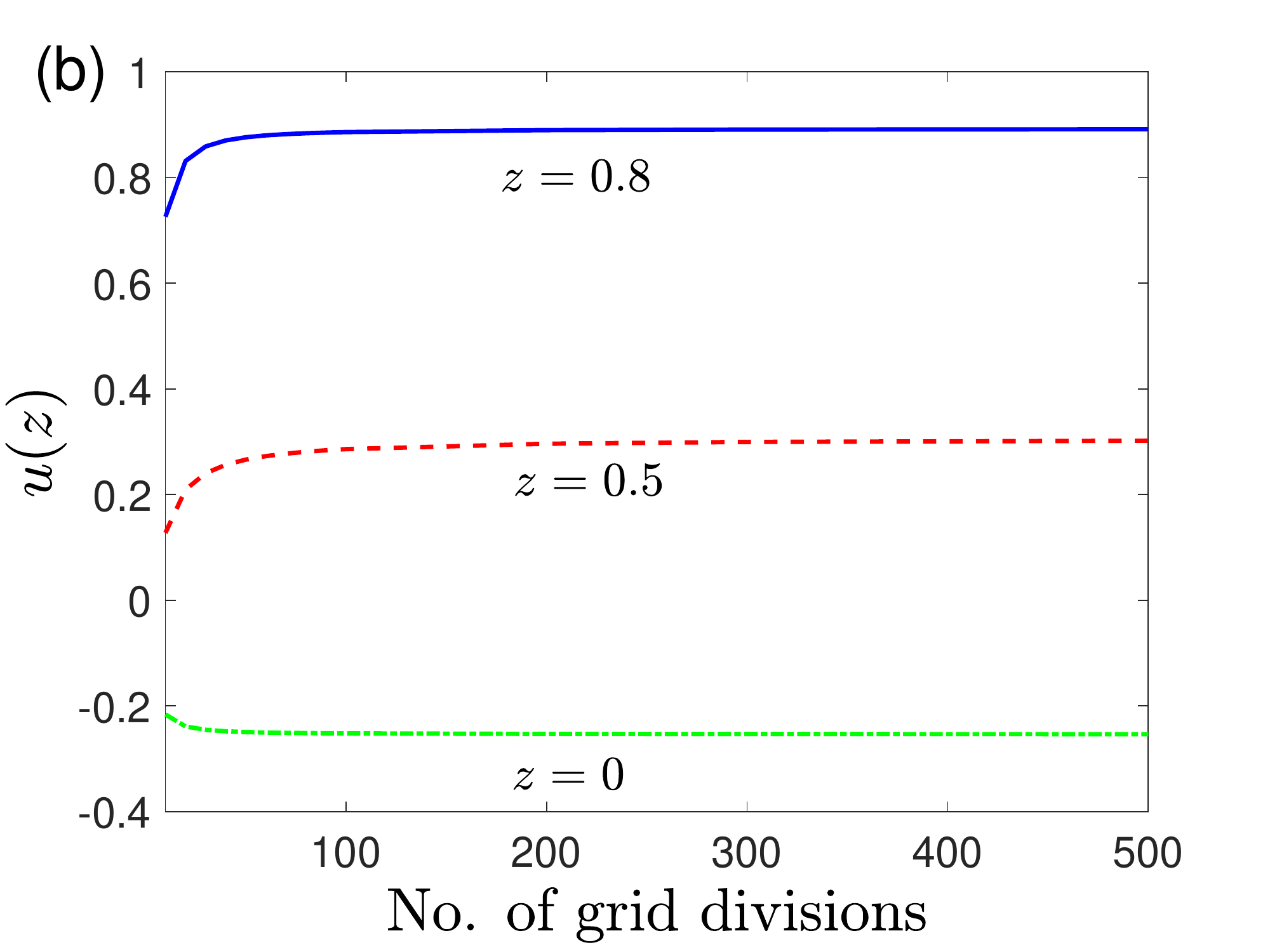}
                \label{fig:k2b}
        \end{subfigure}
 \caption{Grid independence study considering $\alpha = 2.5$, $Re_\Omega = 10$, $\beta = 0.9$, $\kappa=10$ and $d=0.2$  for (a) shear thinning fluid with $n=0.8$ and (b) shear thickening fluid with $n=1.2$. }
\label{fig:k2}
\end{center}
\end{figure}

In order to check our numerical code for consistency, we perform the grid independence study for both shear thinning and shear thickening fluids as shown in figures \ref{fig:k2} (a) and (b) respectively.  At three different locations i.e., $z=0$, $z=0.5$ and $z=0.8$, we plot the change in the value of $u$ as we increase the number of grid divisions. It can be seen from figures \ref{fig:k2} (a) and (b) that going beyond 200 grid points does not change the value of $u$ at any of these locations significantly for both $n=0.8$ as well as $n=1.2$ and therefore we choose the number of grid divisions for our study as 200. Consequently, we get $\Delta z = 0.005$ for our study.

\begin{figure}
\begin{center}
        \begin{subfigure}[t]{0.5\textwidth}
                \includegraphics[width=1\linewidth]{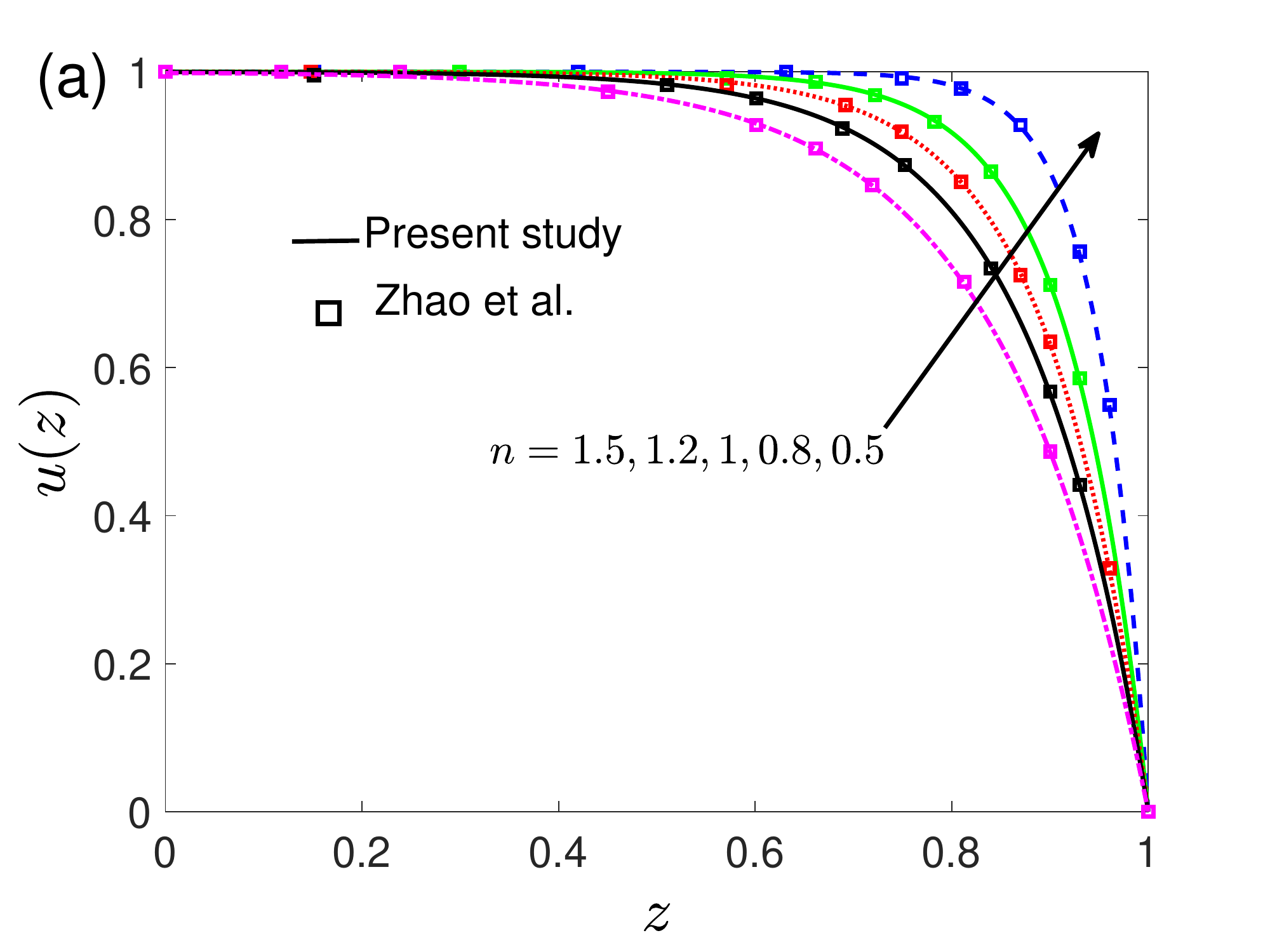}
                \label{fig:k3a}
        \end{subfigure}\hfill
        \begin{subfigure}[t]{0.5\textwidth}
                \includegraphics[width=1\linewidth]{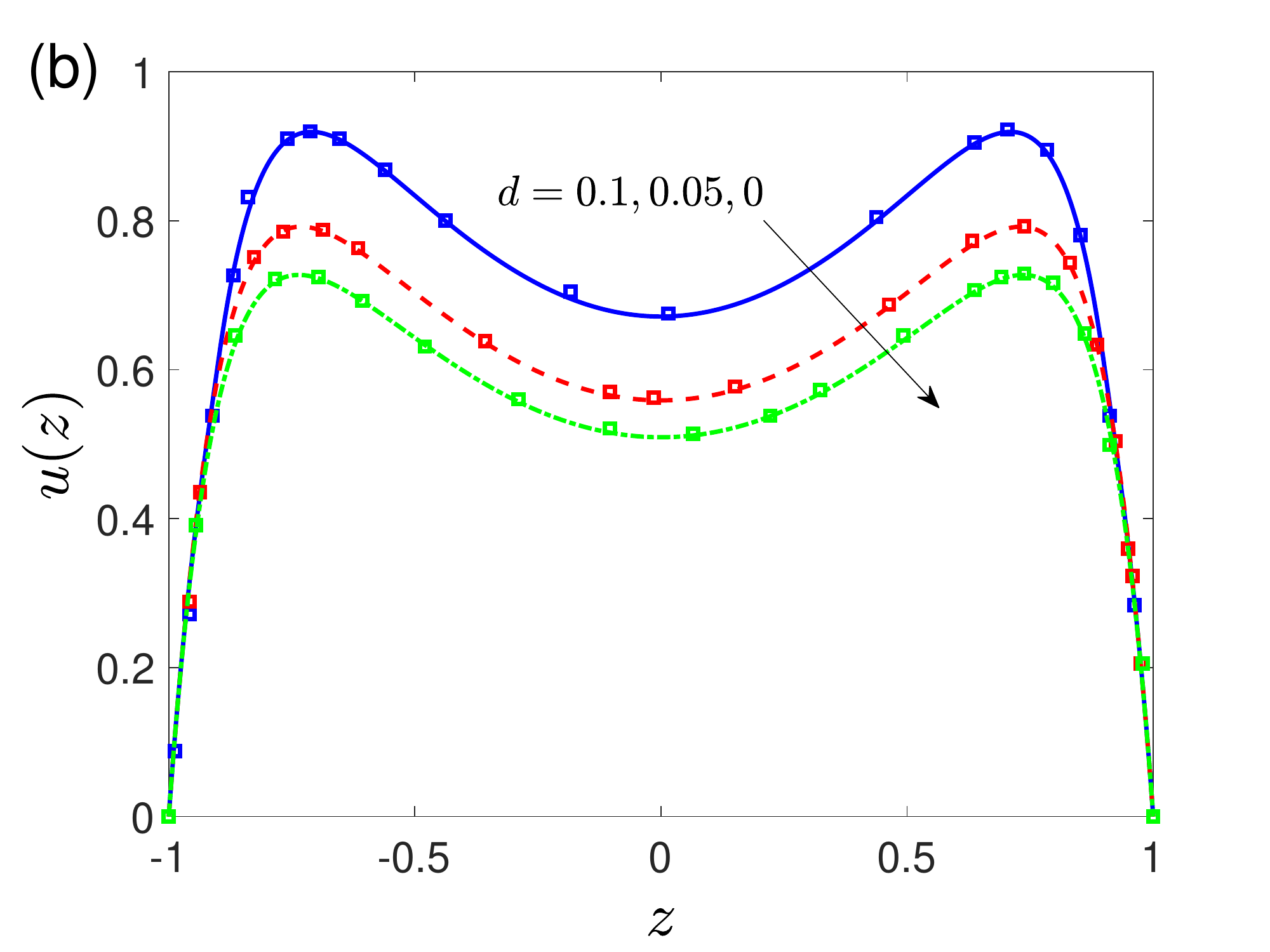}
                \label{fig:k3b}
        \end{subfigure}
 \caption{Comparison of the results from the present study with (a) Zhao et al. \cite{ZHAO2008503} for flow of power-law fluid through slit microchannel with $\kappa = 10$ and (b) with Liu and Jian \cite{Liu2019} for rotational EO flow of Newtonian fluid with $\alpha = 10$, $\kappa=10$ and $\beta=1$  for different PEL thicknes.}
\label{fig:k3}
\end{center}
\end{figure}

Since it is important to validate our code, we compare the results from the present numerical method with the result reported by Zhao et al. \cite{ZHAO2008503}. Zhao et al. \cite{ZHAO2008503} reported on the EOF of power-law fluid in slit microchannel without PEL. We show in figure \ref{fig:k3}(a) the comparison of axial velocity for various values of power law index namely $n=0.5, 0.8, 1, 1.2$ and $1.5$. Other parameters taken for this validation are $\kappa=10$. It is important to mention here that given charge (zeta potential) at the wall was used as boundary condition by Zhao et al. \cite{ZHAO2008503} for the charge distribution equation. Therefore, in order to validate our present numerical code, we use the charge distribution of solution of Zhao et al. \cite{ZHAO2008503} and we find that the present solution shows a good match with the reported results. We also compare our present results with the results of Liu and Jian \cite{Liu2019} for rotational EO flow of Newtonian fluid with different PEL thickness. The comparison is shown in figure \ref{fig:k3}(b). The boundary conditions used to obtain figure \ref{fig:k3}(b) are the same as the ones followed by Liu and Jian \cite{Liu2019} and we see a very good match between our numerical results and the analytical results of the authors. With these two validating studies, it becomes safe to say that our code is suitable for studying the rotational EO flow of power-law fluid throught PE grafted microchannels. 

 The discussion on some of the major results obtained from our work is described in the upcoming section.

\section{Results and discussion}

To effectively understand the flow dynamics of power-law fluid flowing through a PE grafted (soft) microchannel, it is important to understand the role that each parameter plays on the flow. The parameters that affect the flow include, $n$ which represents the effect of fluid rheology, $Re_\Omega$ which represents the rotational speed, $\alpha$ represents PEL drag, $d$ represents PEL thickness, $\beta$ and $\kappa$ represent charge distribution effect. The effect of each of these parameters on flow velocity will be carefully studied and discussed. However, before discussing each of the parameter affecting the flow it is important to understand the range of values of each dimensionless parameter based on the data available in the literature.

\subsection{Parameter selection}

The range of the grafting dimension $d$ is taken to be between $0.01$ and $0.3$ for the present study \cite{chandasinhadas, Liu2019, gaikwad2018softness, kaushikp2019}. The range of dimensionless inverse of EDL thickness $\kappa$ is taken to be of the order of $10$ \cite{stone2004engineering}. The ratio of EDL thickness in the EL to PEL, given by $\beta$ is taken to be between $0.1$ and $1$ \cite{chandasinhadas, Liu2019, gaikwad2018softness, kaushikp2019}. The range of the dimensionless drag parameter $\alpha$ is between $0.1$ and $10$ \cite{chandasinhadas, Liu2019, gaikwad2018softness, kaushikp2019}. The dimensionless rotational speed $Re_\Omega$ is assumed to vary from $0$ to $10$ \cite{changandwang, ng2015electro, QI2017355, kaushikmandalchakraborty2017, ABHIMANYU201656, Kaushik2017, KAUSHIK2017123, kaushikp2019}. The value of power-law index $n$ is assumed to be between $0.6$ and $1.4$. Newtonian fluid is represented by $n=1$. Since the present study is used to understand the flow of bio-fluids such as blood, the value of the power law index commonly used for blood is about $n=0.63$ \cite{MANDAL2007766}.

\subsection{Effect of fluid rheology}
\begin{figure}
\begin{center}
        \begin{subfigure}[t]{0.5\textwidth}
                \includegraphics[width=1\linewidth]{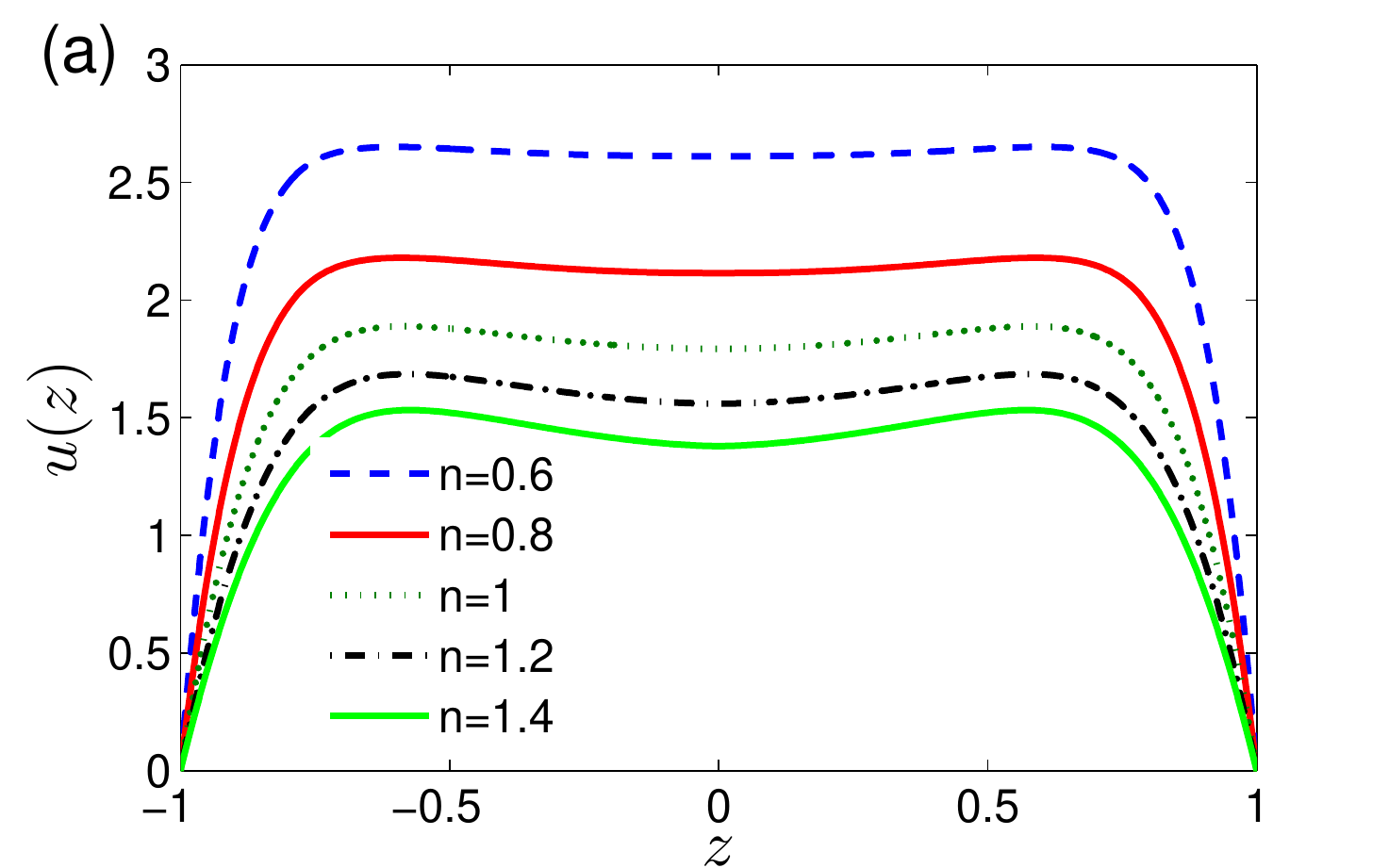}
                \label{fig:k4a}
        \end{subfigure}\hfill
        \begin{subfigure}[t]{0.5\textwidth}
                \includegraphics[width=1\linewidth]{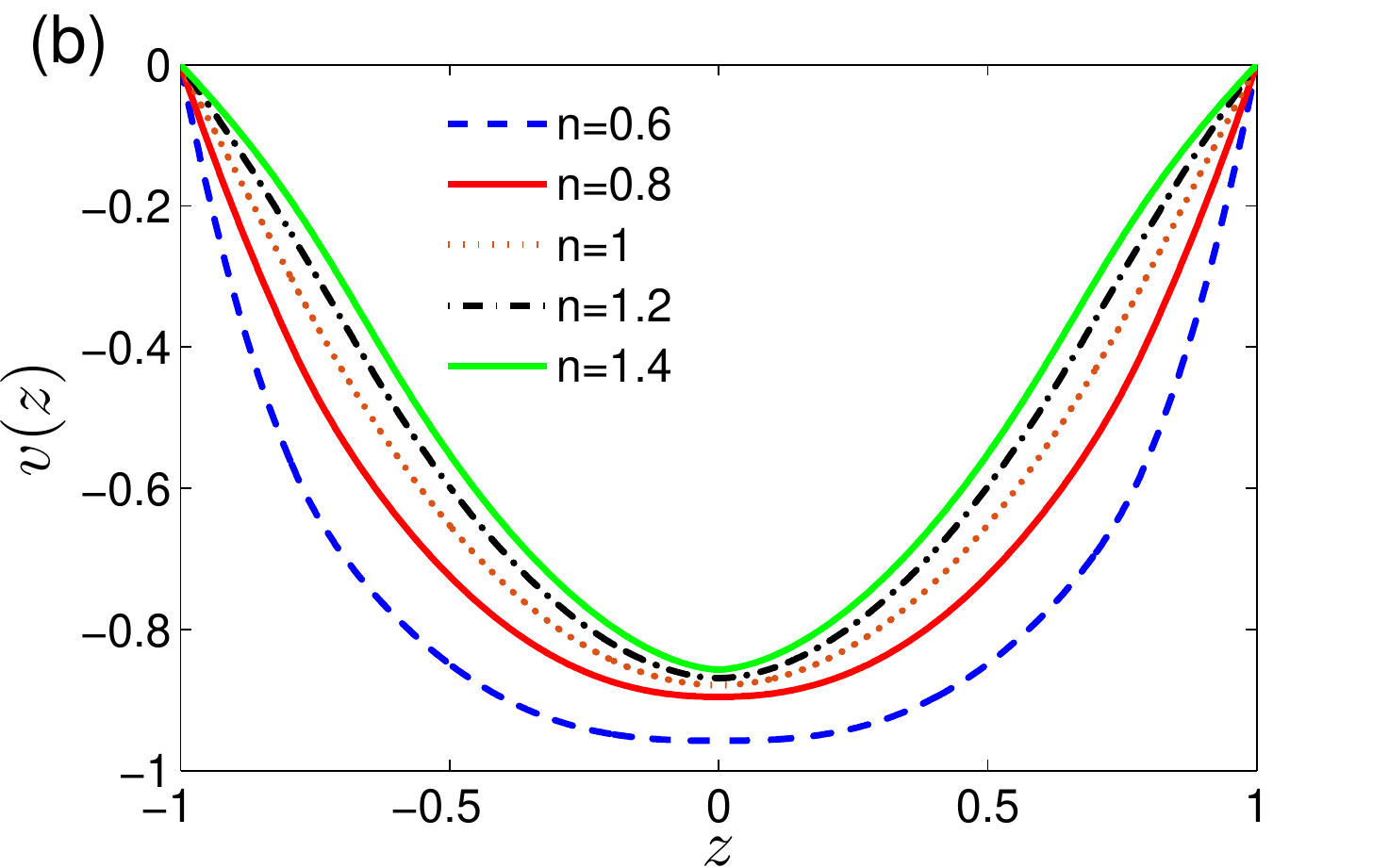}
                \label{fig:k4b}
        \end{subfigure}\\
        \begin{subfigure}[t]{0.5\textwidth}
                \includegraphics[width=1\linewidth]{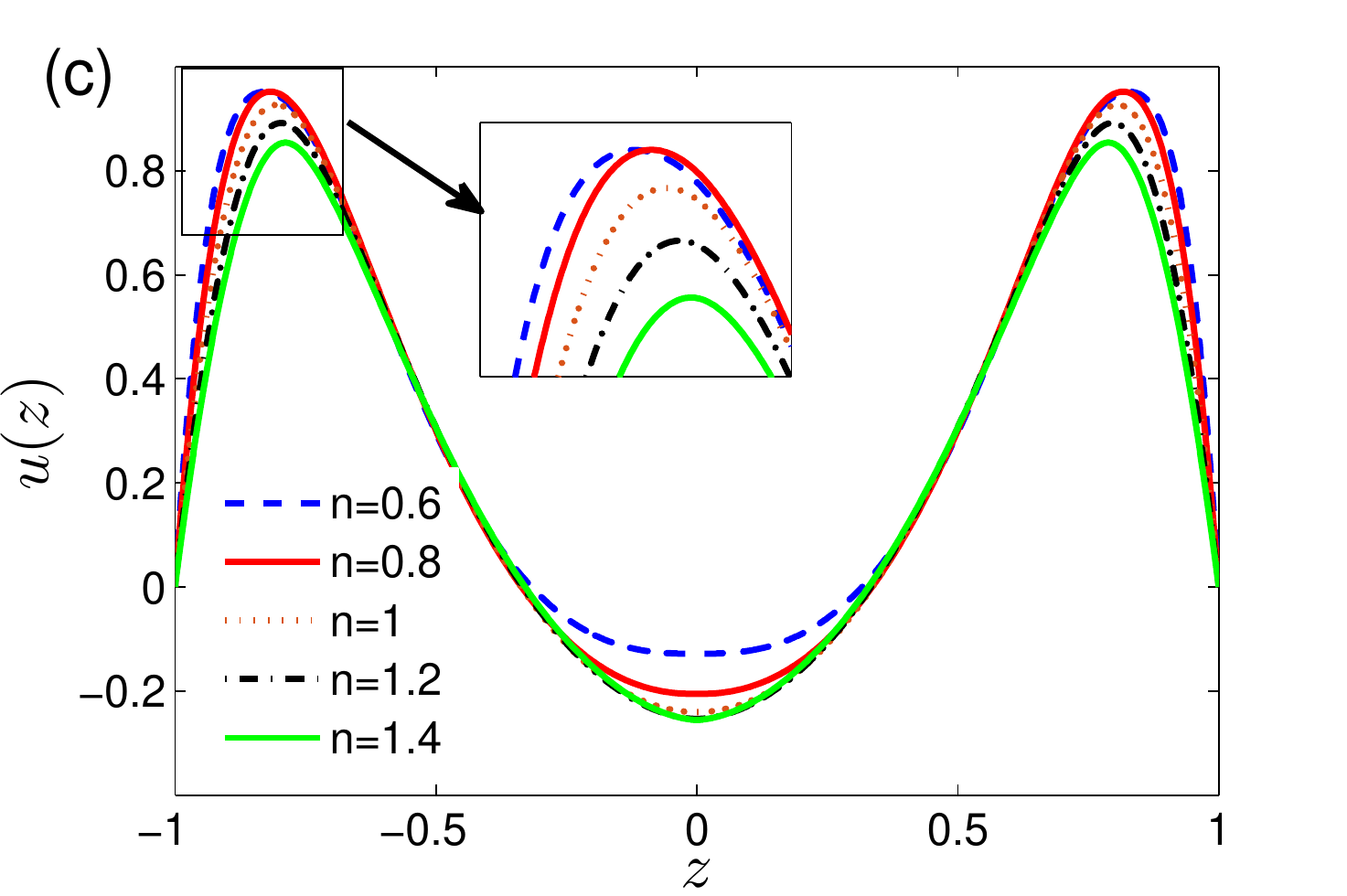}
                \label{fig:k4c}
        \end{subfigure}\hfill
        \begin{subfigure}[t]{0.5\textwidth}
                \includegraphics[width=1\linewidth]{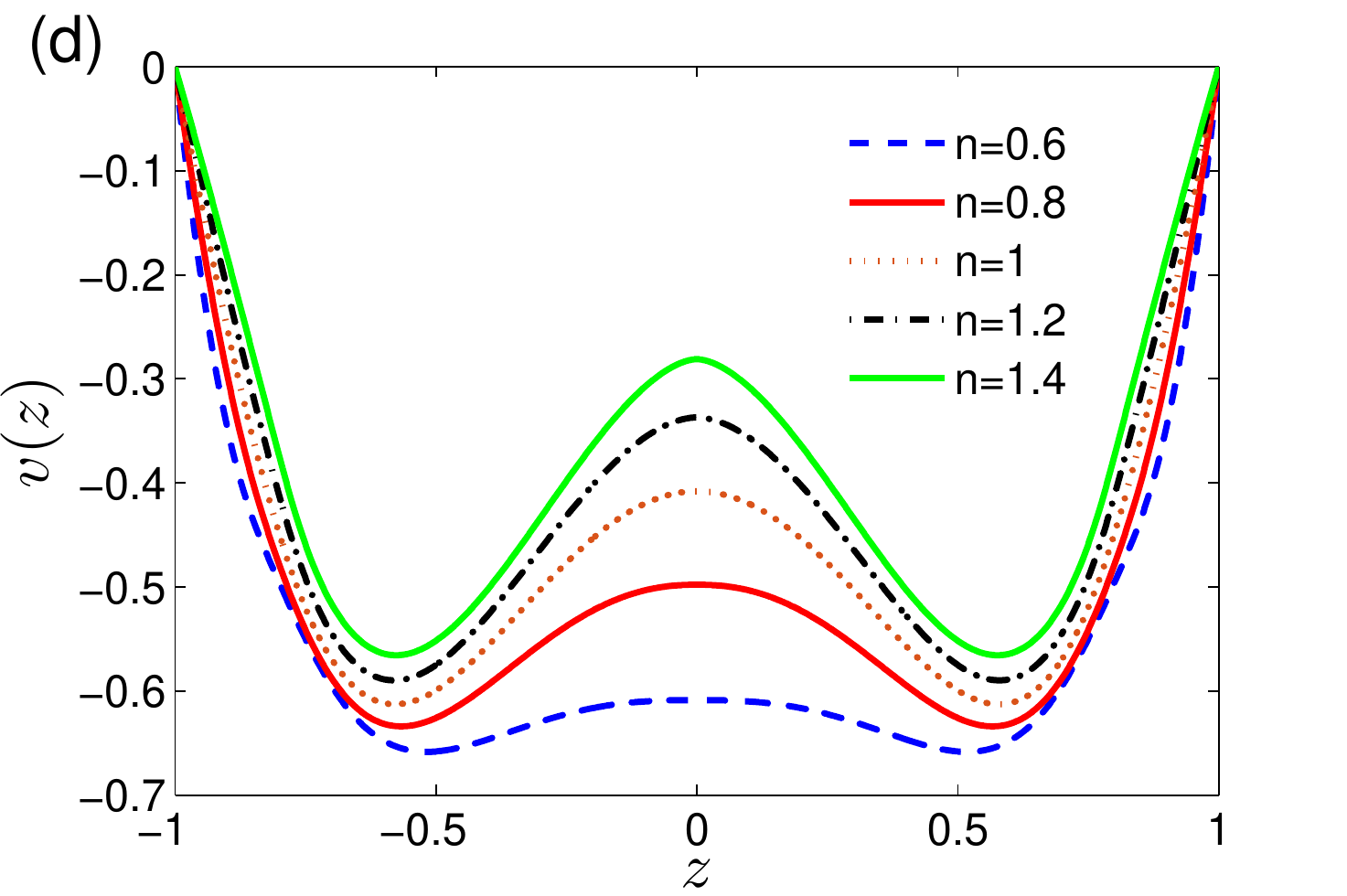}
                \label{fig:k4d}
        \end{subfigure}
 \caption{The variation of (a) $u$-velocity profile for $Re_\Omega= 1$, (b) $v$-velocity profile for $Re_\Omega= 1$, (c) $u$-velocity profile for $Re_\Omega= 10$ and (d) $v$-velocity profile with power-law index $n$ for $Re_\Omega= 10$ with $\alpha = 2.5$, $\kappa=10$, $d=0.2$ and $\beta=0.9$.}
\label{fig:k4}
\end{center}
\end{figure}

The effect of the rheological behaviour of the fluid is understood by varying $n$ - the power law index. The effect of power law index $n$ on the flow velocities at low rotational speeds are depicted in figures \ref{fig:k4} (a), (b) and for high rotational speeds in \ref{fig:k4} (c), (d). It can be seen from the figures that as the value of $n$ increases the velocity magnitude decreases for both $u$ and $v$ because of increase in effective viscosity. We also observe that for low rotational speeds and smaller values of $n$ the velocity profiles are more diffused because of ease of flow of the fluid due to lower effective viscosity. It is also observed from figure \ref{fig:k4}(c) that with for higher rotational speeds the effect of rotation overpowers the effect of fluid rheology, thereby making the distinction between the different $u$ velocity profiles very small. However, at high rotational speeds, the distinction in the $v$ velocity profiles is still clear as seen from figure \ref{fig:k4}(d). The $v$ velocity has a reduction in magnitude close to the centre of the channel consistent with the results of Kaushik et al. \cite{kaushikp2019} as well as Liu and Jian \cite{Liu2019}. It is also important to note that the variation in $v$ velocity is higher for higher values of $n$ in figure \ref{fig:k4}(d) since for lower values of $n$, effective viscosity is lower and therefore external effects are not felt in a very pronouced way.

\subsection{Effect of fluid rotational speed}
\begin{figure}
\begin{center}
        \begin{subfigure}[t]{0.5\textwidth}
                \includegraphics[width=1\linewidth]{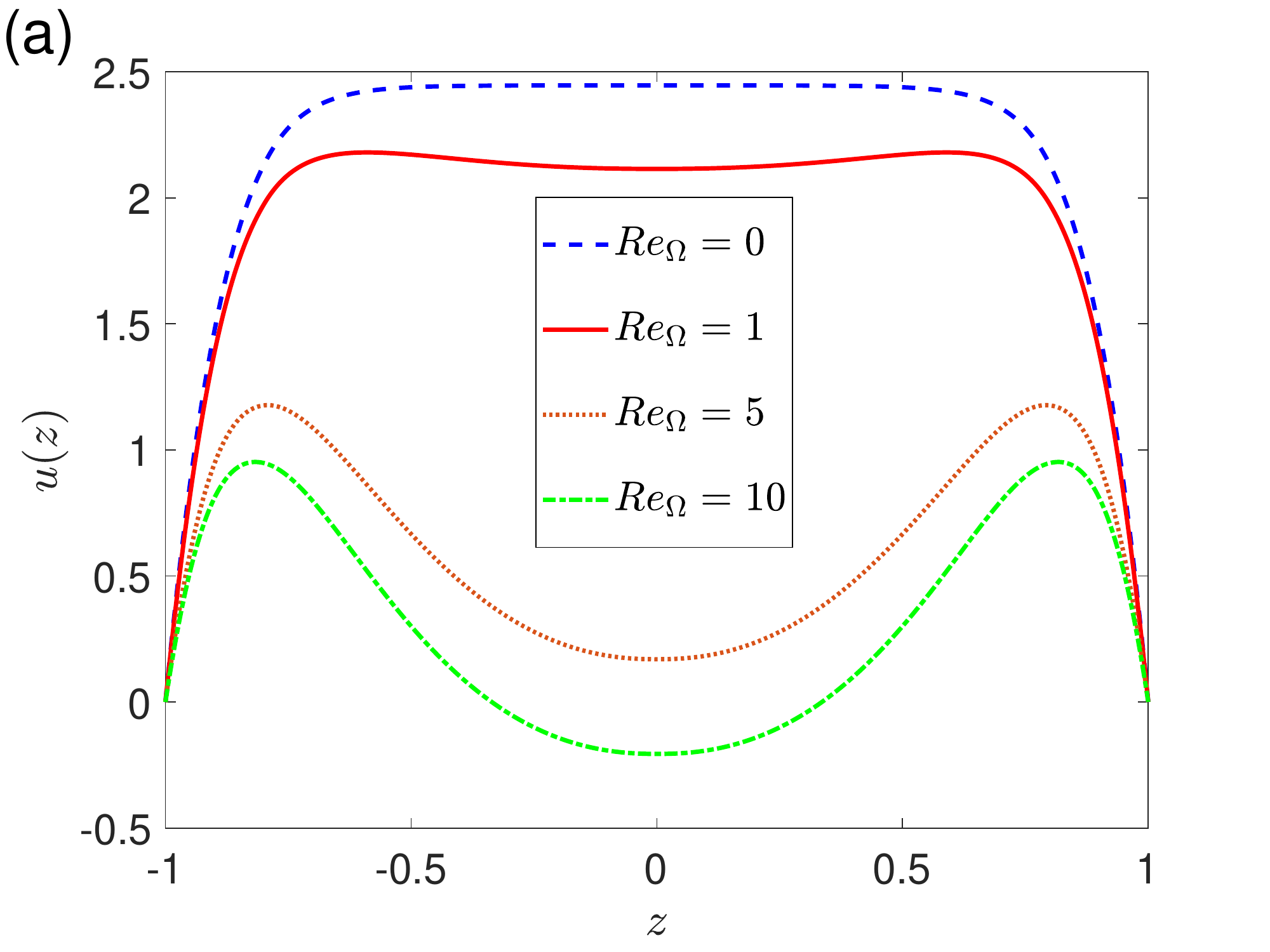}
                \label{fig:k5a}
        \end{subfigure}\hfill
        \begin{subfigure}[t]{0.5\textwidth}
                \includegraphics[width=1\linewidth]{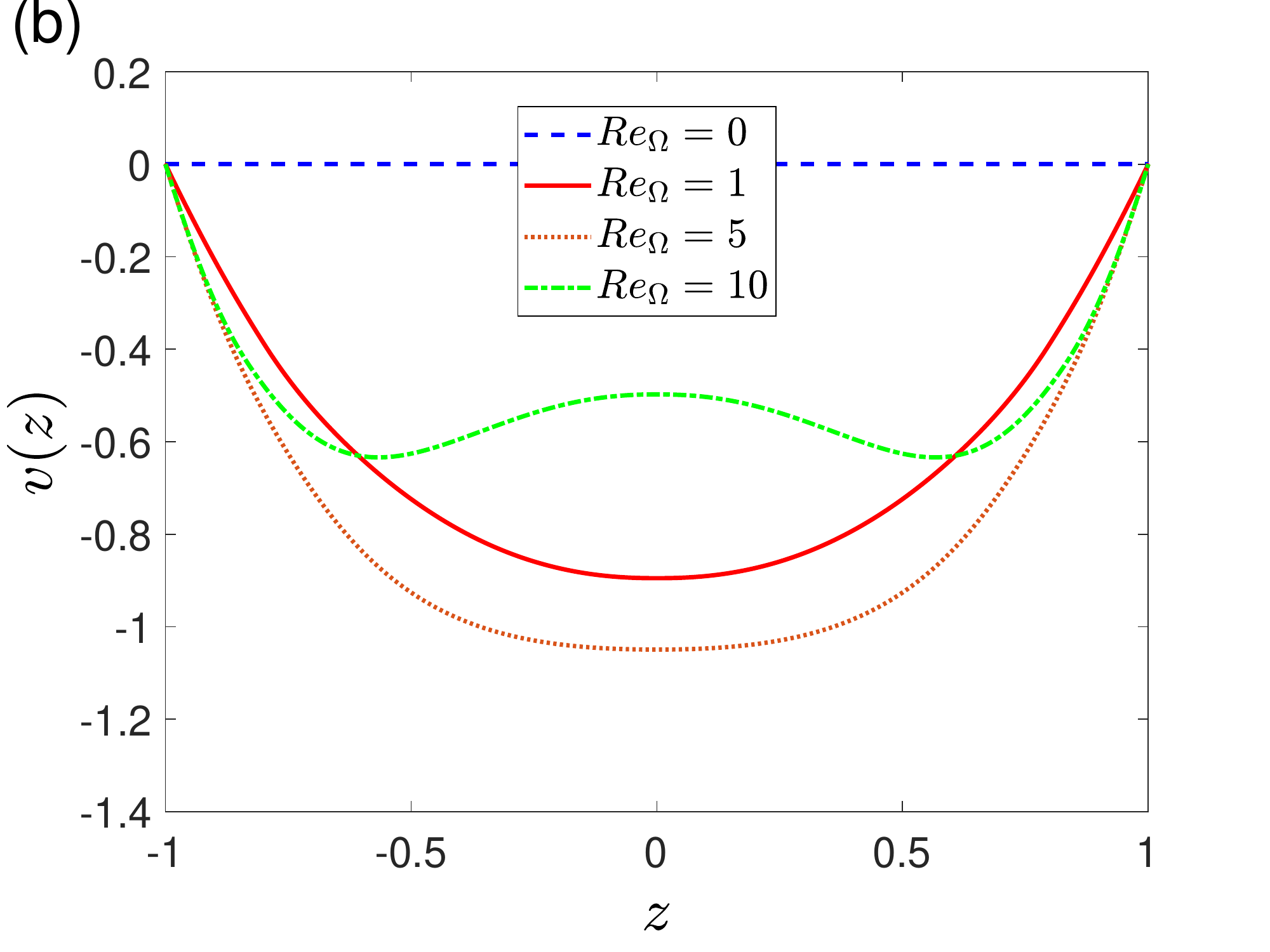}
                \label{fig:k5b}
        \end{subfigure}\\
        \begin{subfigure}[t]{0.5\textwidth}
                \includegraphics[width=1\linewidth]{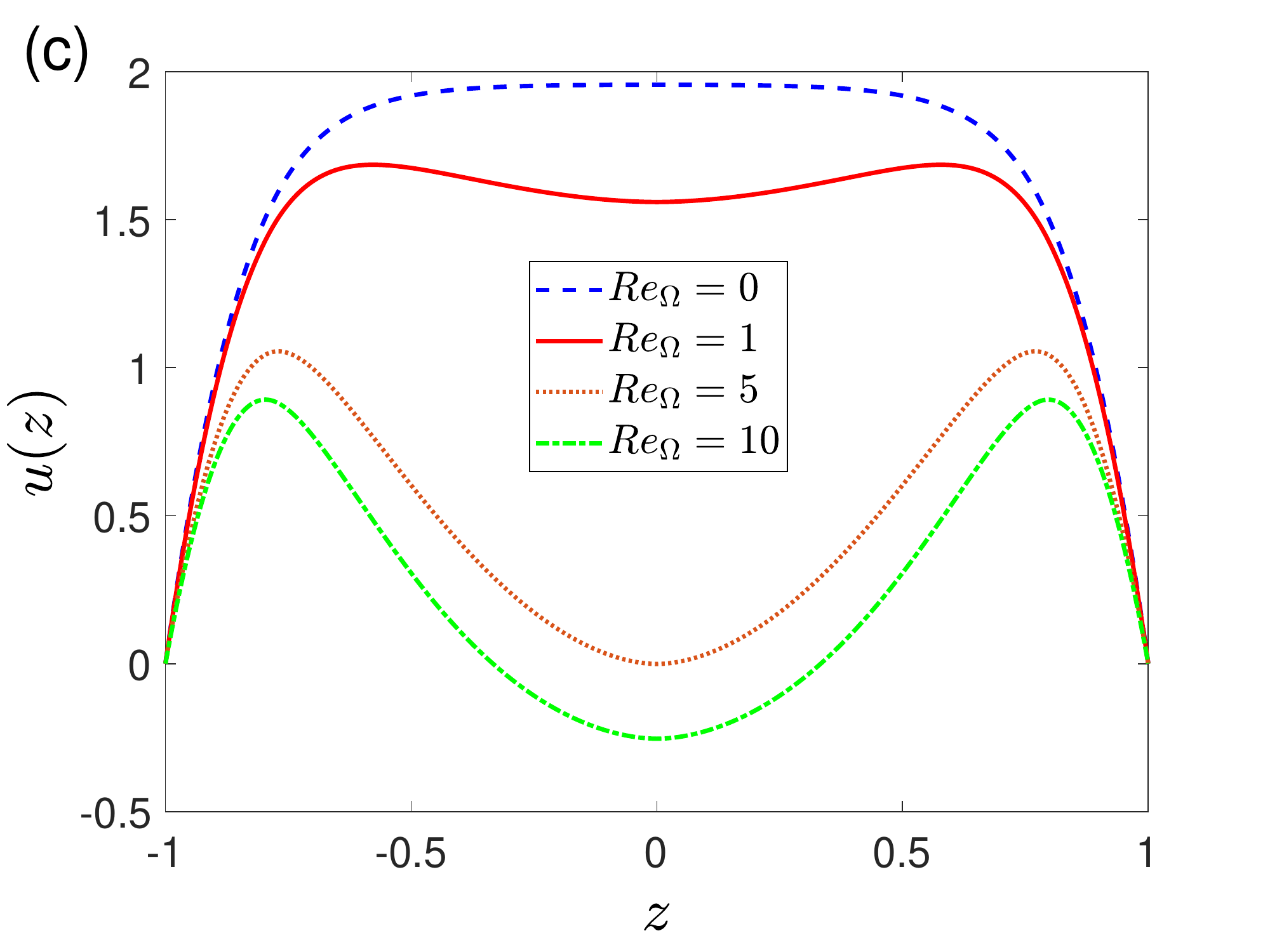}
                \label{fig:k5c}
        \end{subfigure}\hfill
        \begin{subfigure}[t]{0.5\textwidth}
                \includegraphics[width=1\linewidth]{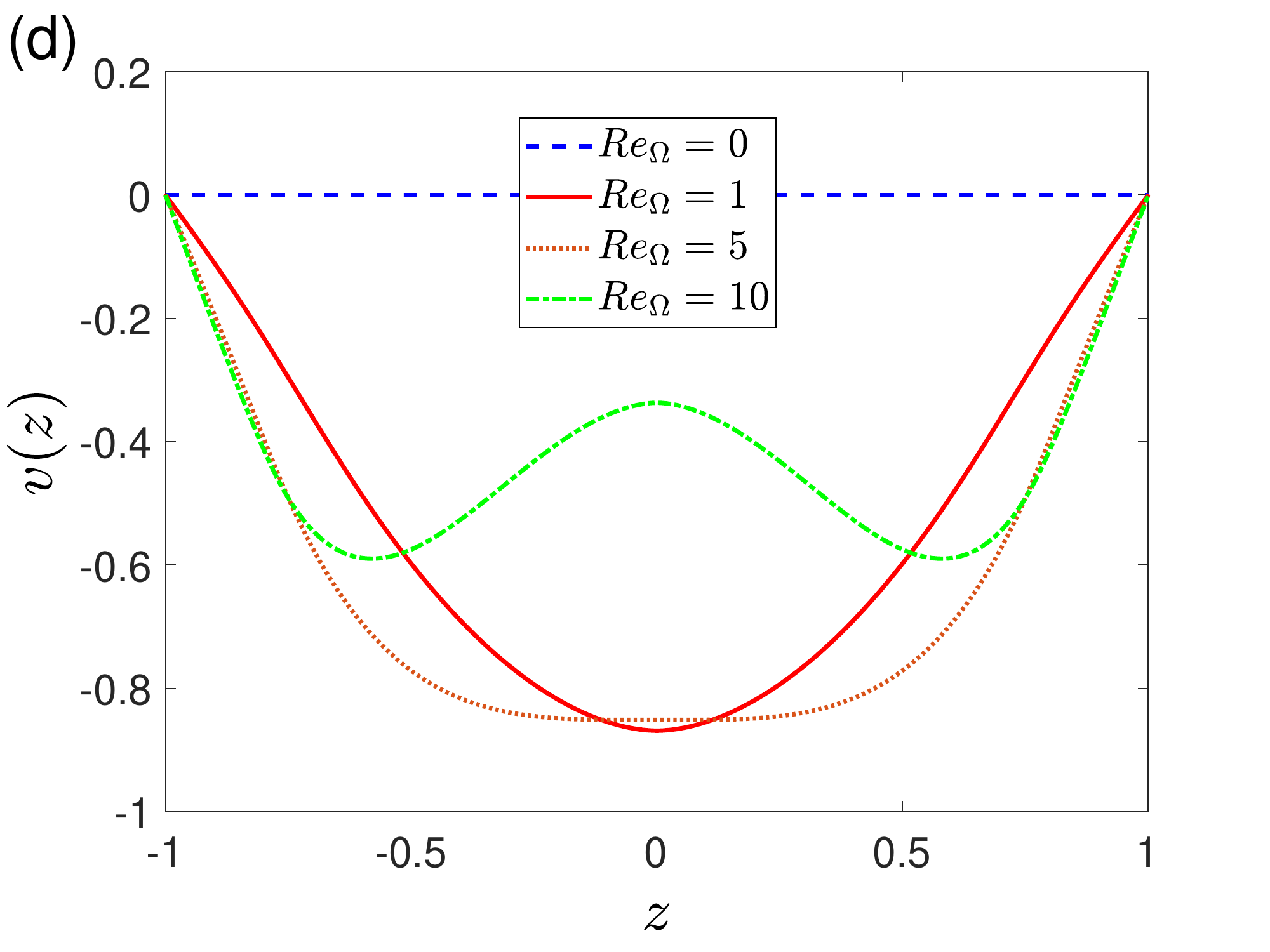}
                \label{fig:k5d}
        \end{subfigure}
 \caption{The variation of (a) $u$-velocity profile for $n= 0.8$, (b) $v$-velocity profile for $n= 0.8$, (c) $u$-velocity profile for $n= 1.2$ and (d) $v$-velocity profile  for $n= 1.2$ with different dimensionless rotational speed $Re_\Omega$ with $\alpha = 2.5$, $\kappa=10$, $d=0.2$ and $\beta=0.9$.}
\label{fig:k5}
\end{center}
\end{figure}

It is important to understand the Coriolis force based alterations on the fluid flow and therefore in figure \ref{fig:k5} we plot the variation of $u$ and $v$ velocity profiles when $Re_\Omega$ is varied for both shear thinning and shear thickening fluids. With increase in $Re_\Omega$, there is decrease in $u$ velocity for both shear thinning and shear thickening fluids as seen from figures \ref{fig:k5} (a) and (c). This decrease is because of transfer of axial flow energy to drive the transverse direction flow as the rotational speed increases. This observation is consistent with results reported by Chang and Wang \cite{changandwang}. It also important to observe that irrespective of the power-law index, the overall dimensionless velocity magnitude is higher when PEL exists as compared to rigid channels having rotational EO flow studied by Chang and Wang \cite{changandwang}. One may observe that as $Re_\Omega$  increases the $v$ velocity first increases in magnitude and then decreases as seen from figures \ref{fig:k5} (b) and (d). This is because of two-way coupling of the $u$ and $v$ velocities owing to the Coriolis force term in the momentum equations. Initially as $u$ magnitude decreases with rotation, $v$ magnitude increases to maintain the loss of energy from $x$-direction momentum. As $v$ magnitude further goes on increasing, it has an effect on the Coriolis term in the $x$-direction momentum equation and causes $u$ magnitude to decrease further and thereby decreasing the $v$ magnitude as well. This can be seen in \ref{fig:k5} (b), where $v$ velocity magnitude increases when $Re_\Omega$ is increased from $1$ to $5$ and further increase to $10$ decreases the $v$ velocity magnitude.

\subsection{Effect of PEL}
\begin{figure}
\begin{center}
        \begin{subfigure}[t]{0.5\textwidth}
                \includegraphics[width=1\linewidth]{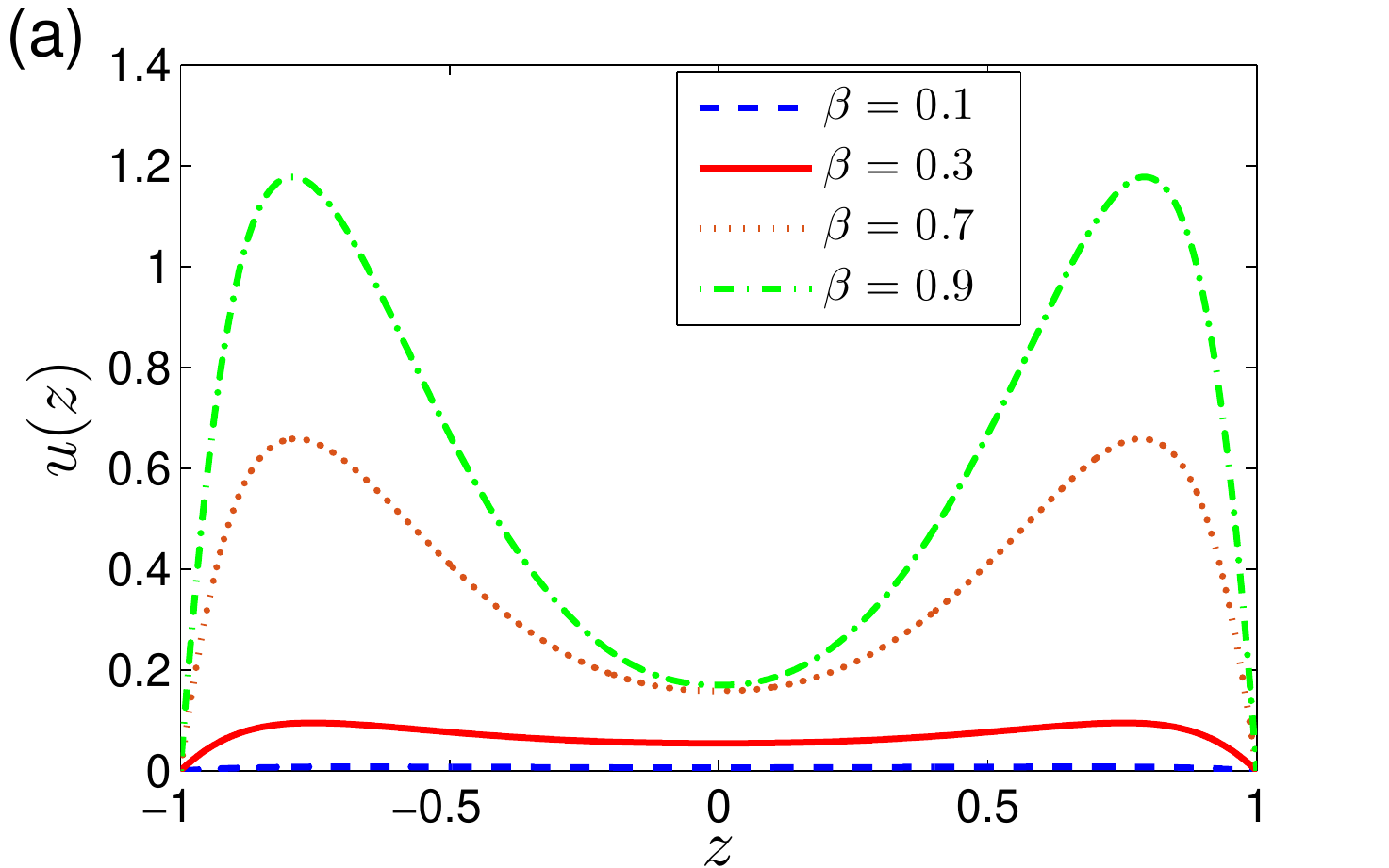}
                \label{fig:k6a}
        \end{subfigure}\hfill
        \begin{subfigure}[t]{0.5\textwidth}
                \includegraphics[width=1\linewidth]{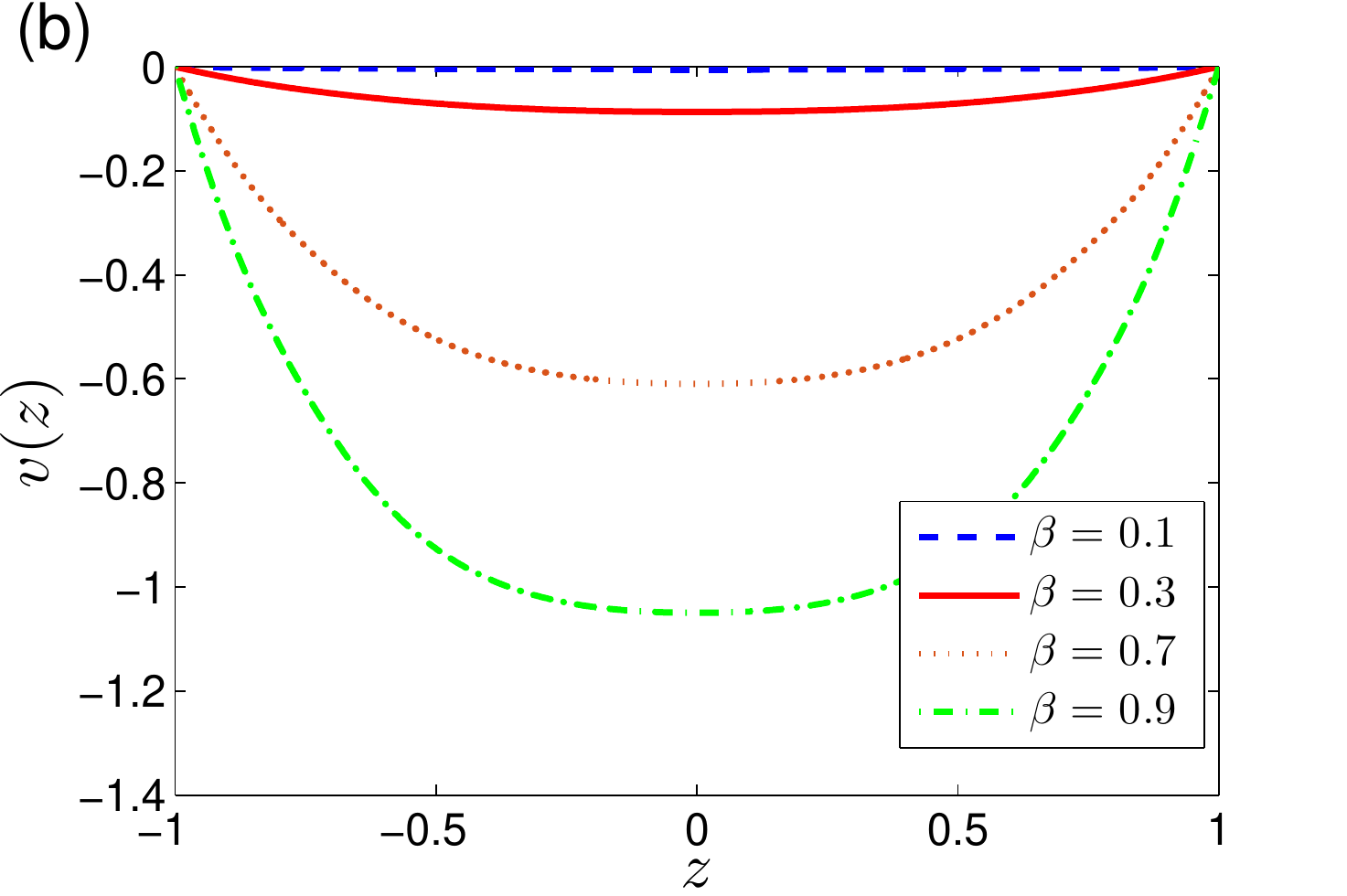}
                \label{fig:k6b}
        \end{subfigure}\\
        \begin{subfigure}[t]{0.5\textwidth}
                \includegraphics[width=1\linewidth]{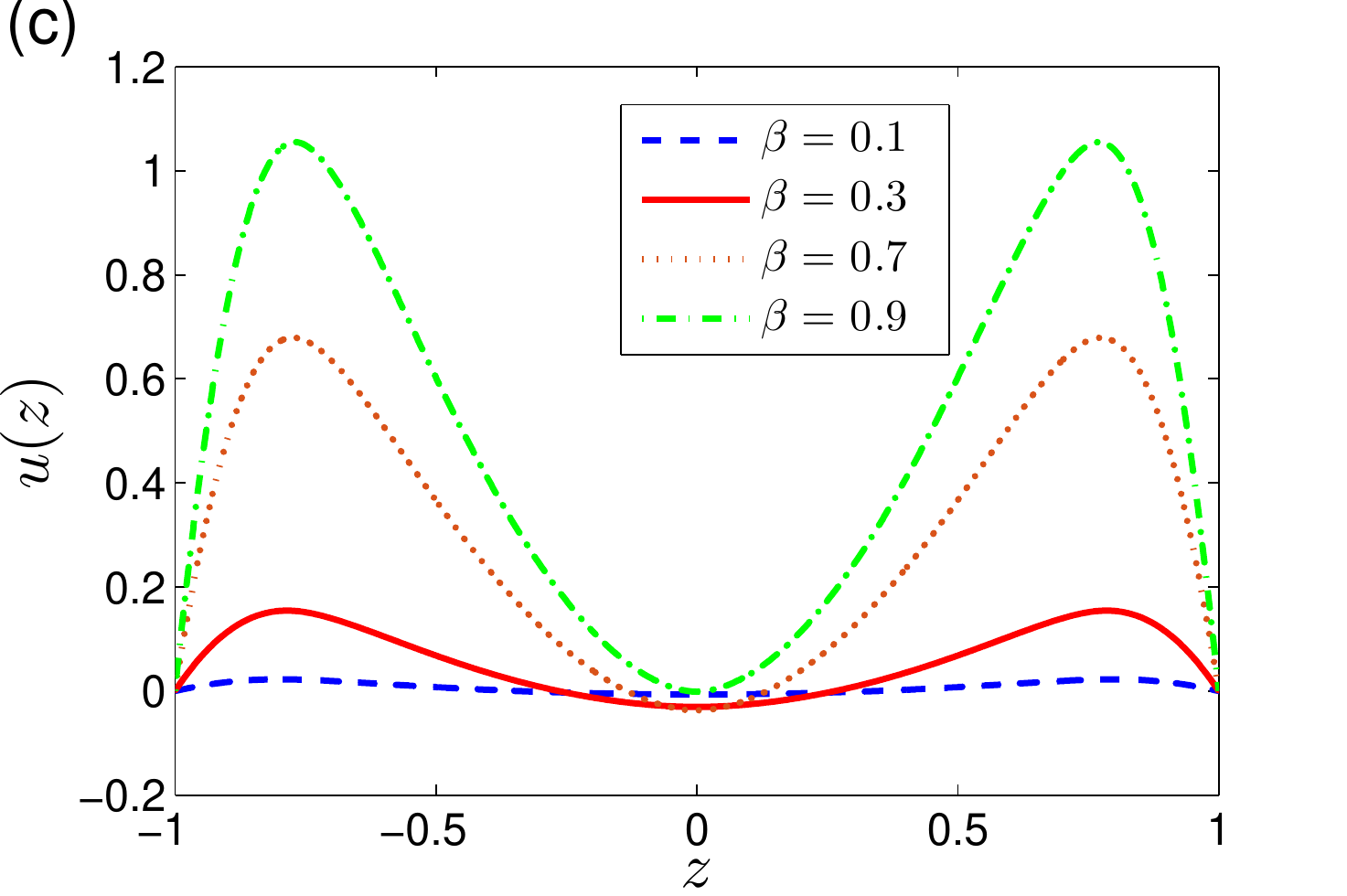}
                \label{fig:k6c}
        \end{subfigure}\hfill
        \begin{subfigure}[t]{0.5\textwidth}
                \includegraphics[width=1\linewidth]{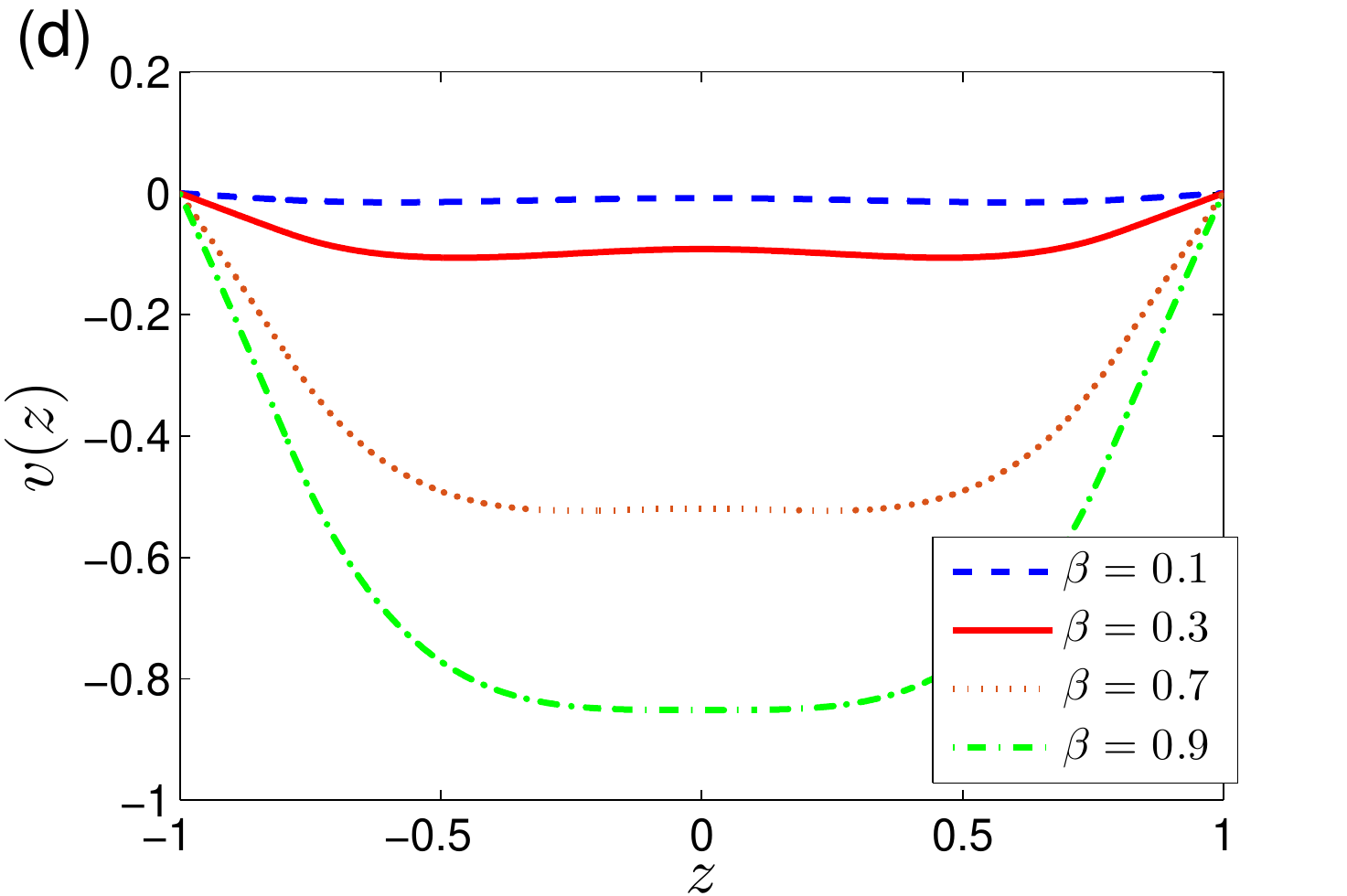}
                \label{fig:k6d}
        \end{subfigure}
 \caption{The variation of (a) $u$-velocity profile for $n= 0.8$, (b) $v$-velocity profile for $n= 0.8$, (c) $u$-velocity profile for $n= 1.2$ and (d) $v$-velocity profile  for $n= 1.2$ for different dimensionless inverse of EDL thickness in the PEL $\beta$ with $Re_\Omega=5$, $\alpha = 2.5$, $\kappa=10$ and $d=0.2$}
\label{fig:k6}
\end{center}
\end{figure}

The effect of PEL on the flow may be examined by examining the effect of EDL thickness inside the PEL. Figure \ref{fig:k6} shows the variation of flow velocities for shear thinning and shear thickening fluids with different values of $\beta$. As $\beta$ decreases, the thickness of the EDL inside the PEL decreases and we see a drastic increasing in both $u$ and $v$ velocities for both shear thinning and shear thickening fluids. It can be seen from figure \ref{fig:k6} that $\beta$ has a very strong impact on the flow velocites irrespective of the value of $n$. It can also be seen from figures \ref{fig:k6} (a) and (c) that decrease in the value of $\beta$ tends to make the $u$ velocity magnitude closer to the walls higher. This is because with decrease in $\beta$ the value of $\psi$ at the wall shows a very large increase for the Gaussian boundary condition considered in the present study. This trend is consistent with data obtained by Chanda et al. \cite{chandasinhadas} and Kaushik et al. \cite{kaushikp2019}.

\begin{figure}
\begin{center}
        \begin{subfigure}[t]{0.5\textwidth}
                \includegraphics[width=1\linewidth]{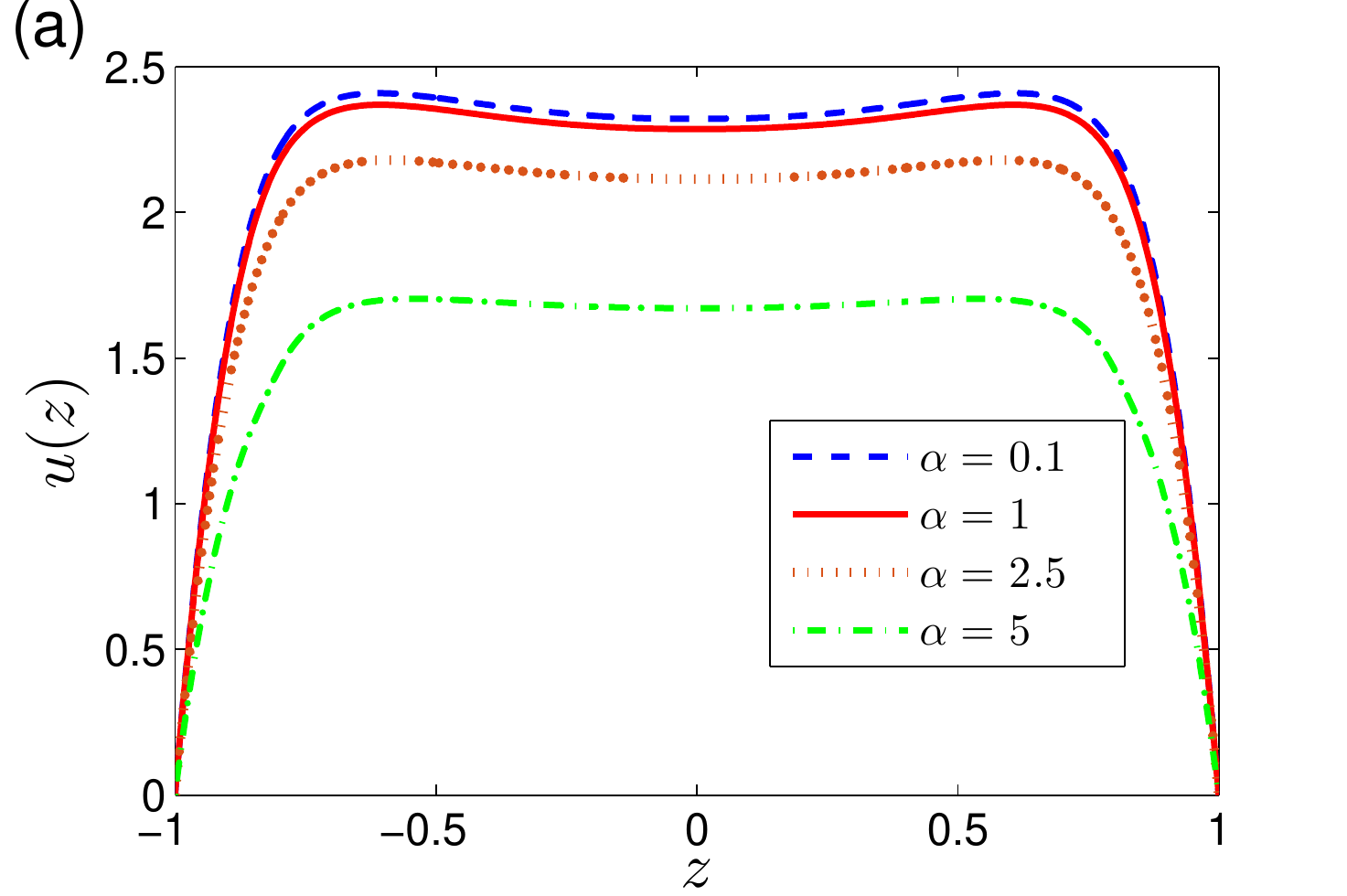}
                \label{fig:k7a}
        \end{subfigure}\hfill
        \begin{subfigure}[t]{0.5\textwidth}
                \includegraphics[width=1\linewidth]{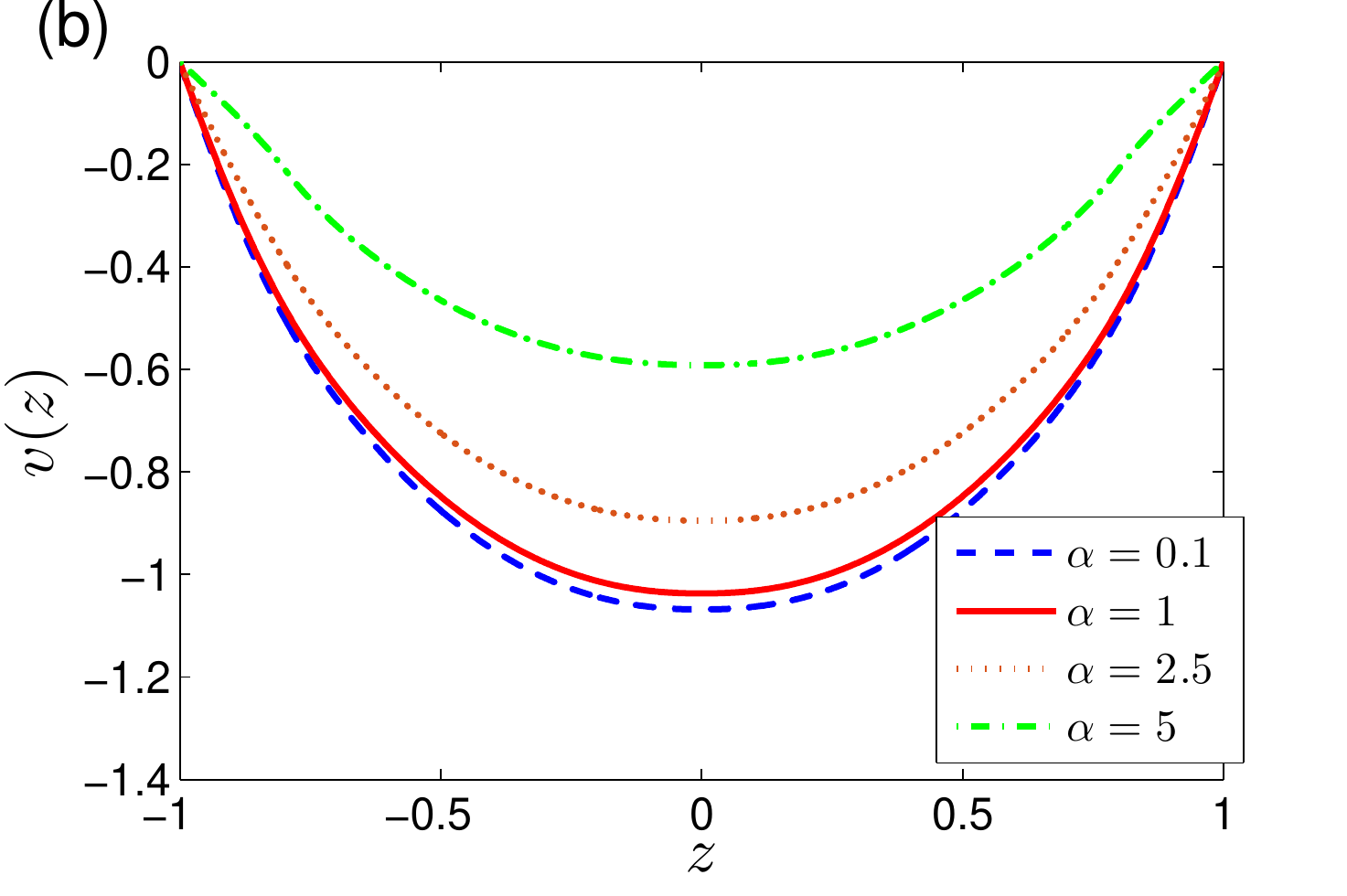}
                \label{fig:k7b}
        \end{subfigure}\\
        \begin{subfigure}[t]{0.5\textwidth}
                \includegraphics[width=1\linewidth]{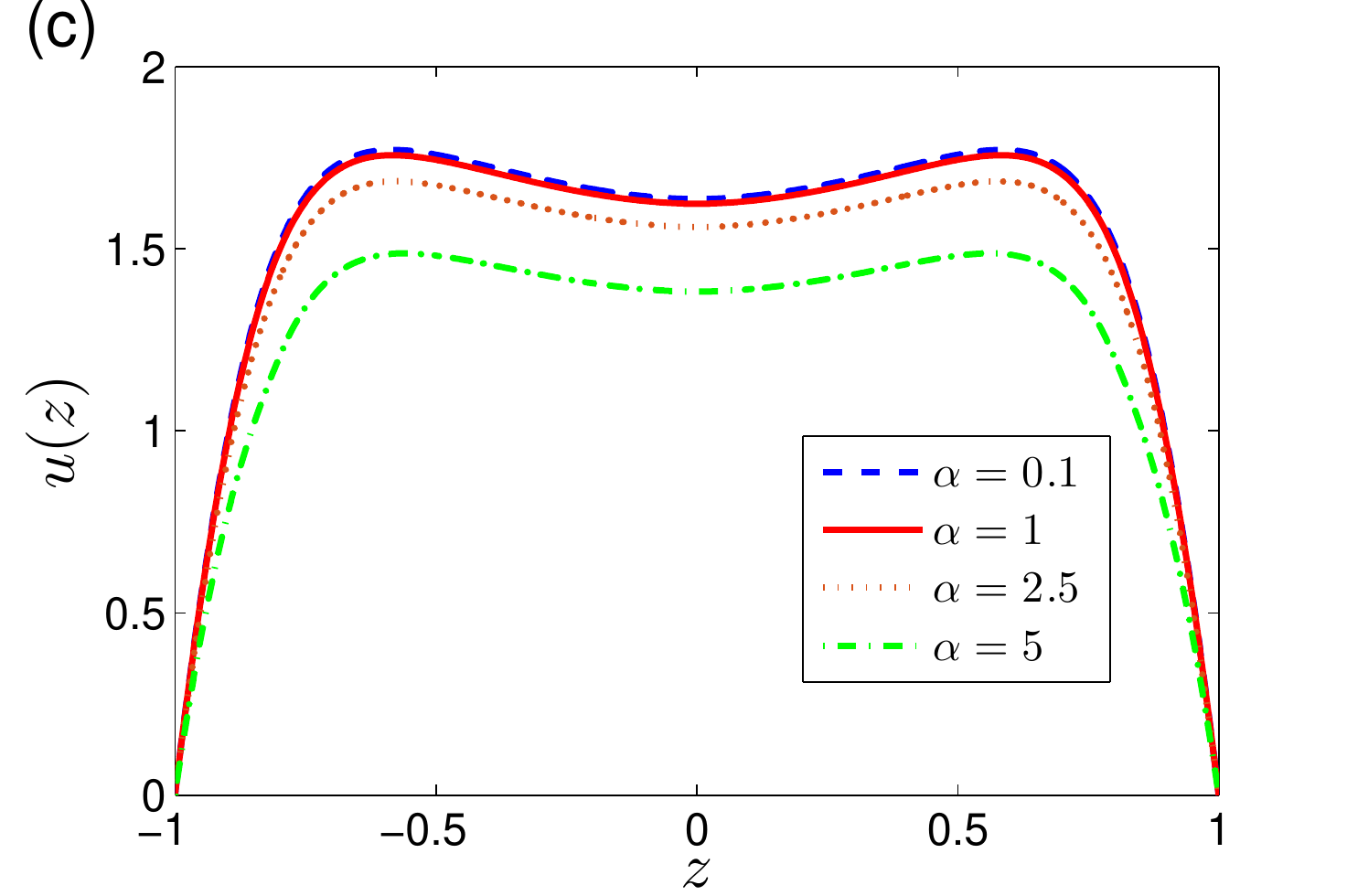}
                \label{fig:k7c}
        \end{subfigure}\hfill
        \begin{subfigure}[t]{0.5\textwidth}
                \includegraphics[width=1\linewidth]{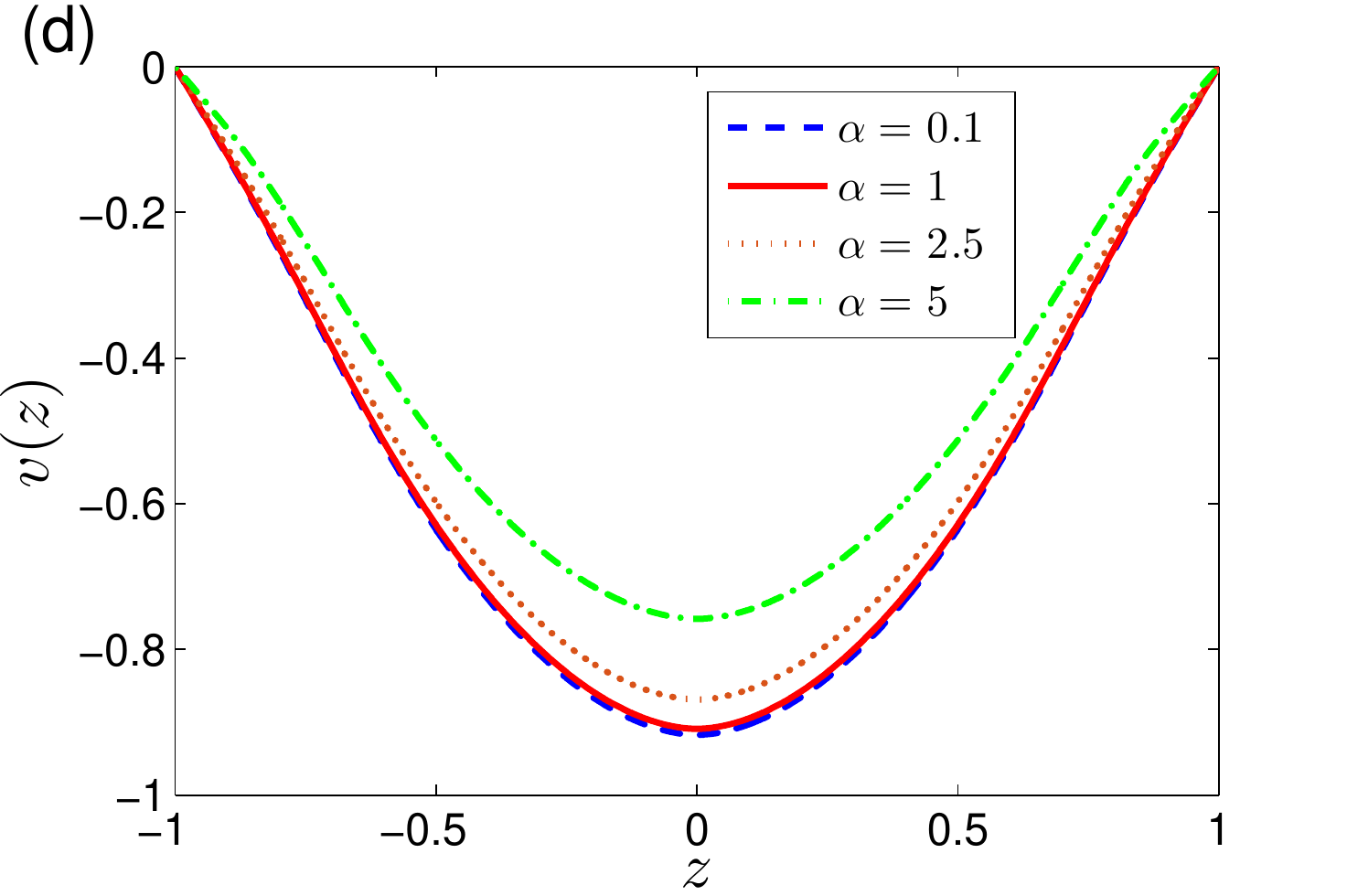}
                \label{fig:k7d}
        \end{subfigure}
 \caption{The variation of (a) $u$-velocity profile for $n= 0.8$, (b) $v$-velocity profile for $n= 0.8$, (c) $u$-velocity profile for $n= 1.2$ and (d) $v$-velocity profile  for $n= 1.2$ for different dimensionless PEL drag $\alpha$ with $Re_\Omega=5$, $\beta = 0.9$, $\kappa=10$ and $d=0.2$}
\label{fig:k7}
\end{center}
\end{figure}

It is also important to discuss the effect of drag experienced by the fluid in the PEL. This effect can be studied by varying $\alpha$. As $\alpha$ increases the drag increases. The effect of changing $\alpha$ on the flow velocities is shown in figure \ref{fig:k7}. It can be seen from figures \ref{fig:k7} (a), (b), (c) and (d) that as we increase the value of $\alpha$ the velocity magnitude decreases for both $u$ and $v$ velocity. The effect of decrease in the velocities is more pronounced for shear thinning fluids as compared to shear thickening fluids as seen from figure \ref{fig:k7}. Changing $\alpha$ only changes the magnitude of the flow velocites and does not have any significant effect on the shape of the velocity profile. This observation is consistent with observation of Kaushik et al. \cite{kaushikp2019} as well as Liu and Jian \cite{Liu2019} for Newtonian fluids. 

\begin{figure}
\begin{center}
        \begin{subfigure}[t]{0.5\textwidth}
                \includegraphics[width=1\linewidth]{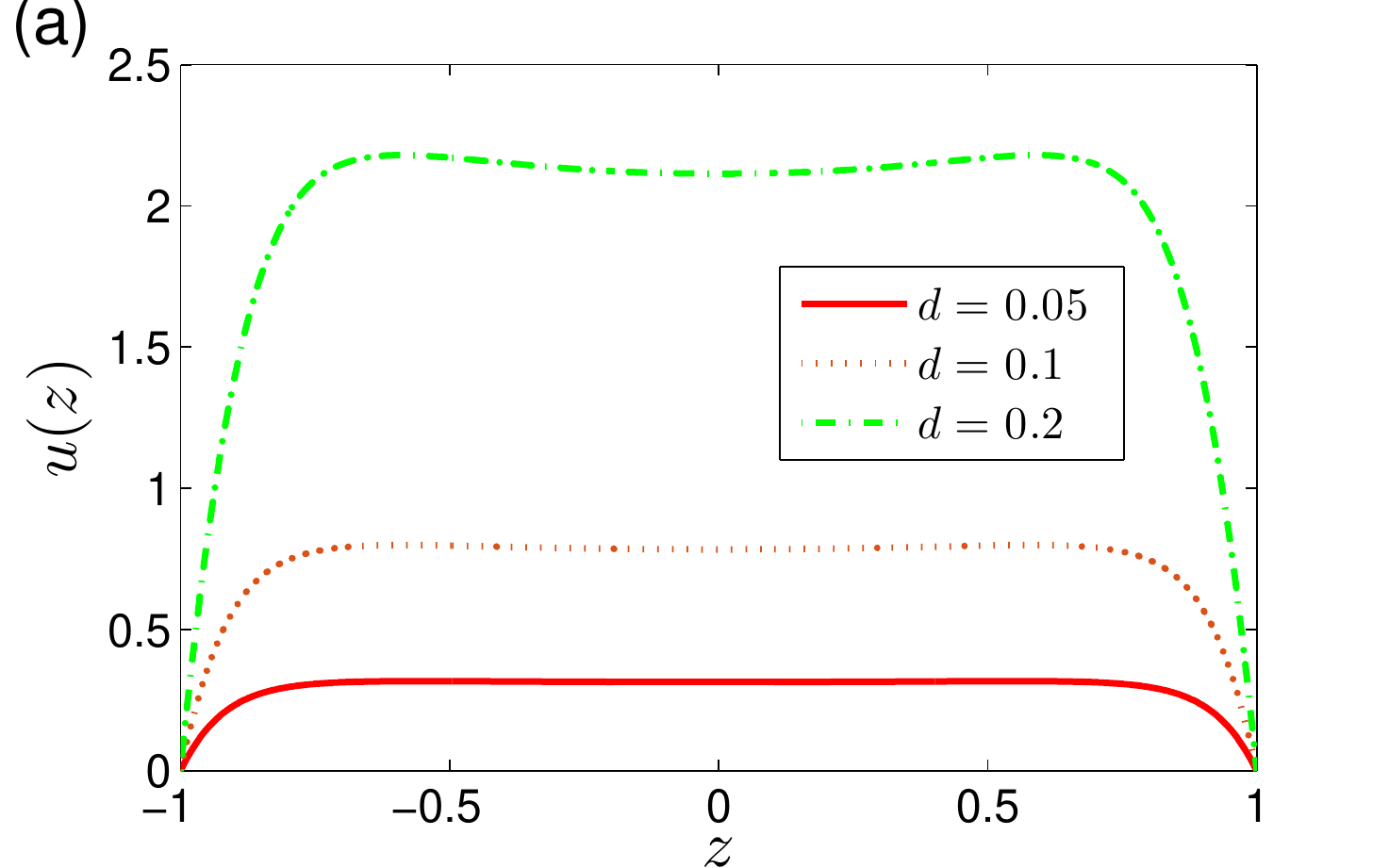}
                \label{fig:k8a}
        \end{subfigure}\hfill
        \begin{subfigure}[t]{0.5\textwidth}
                \includegraphics[width=1\linewidth]{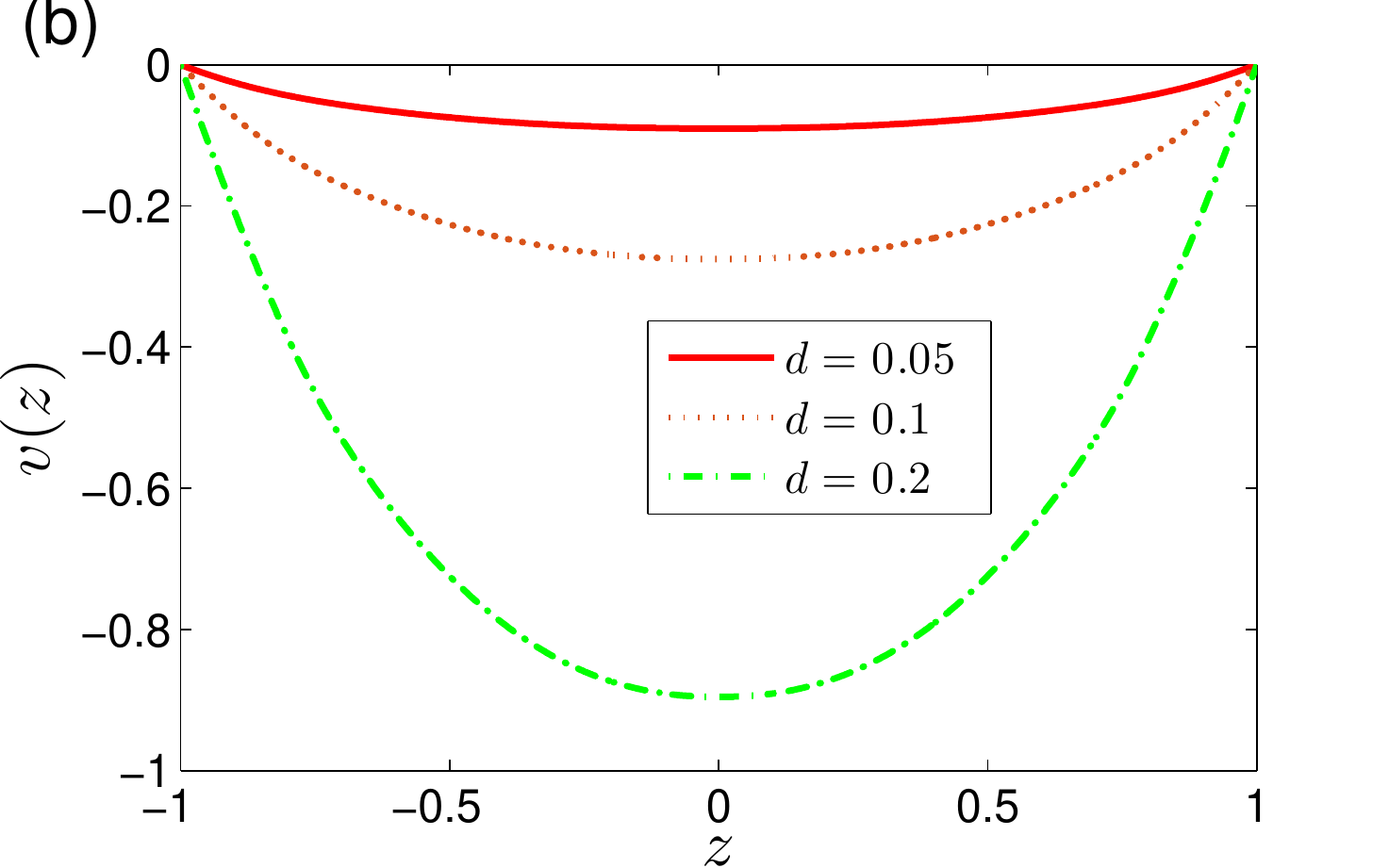}
                \label{fig:k8b}
        \end{subfigure}\\
        \begin{subfigure}[t]{0.5\textwidth}
                \includegraphics[width=1\linewidth]{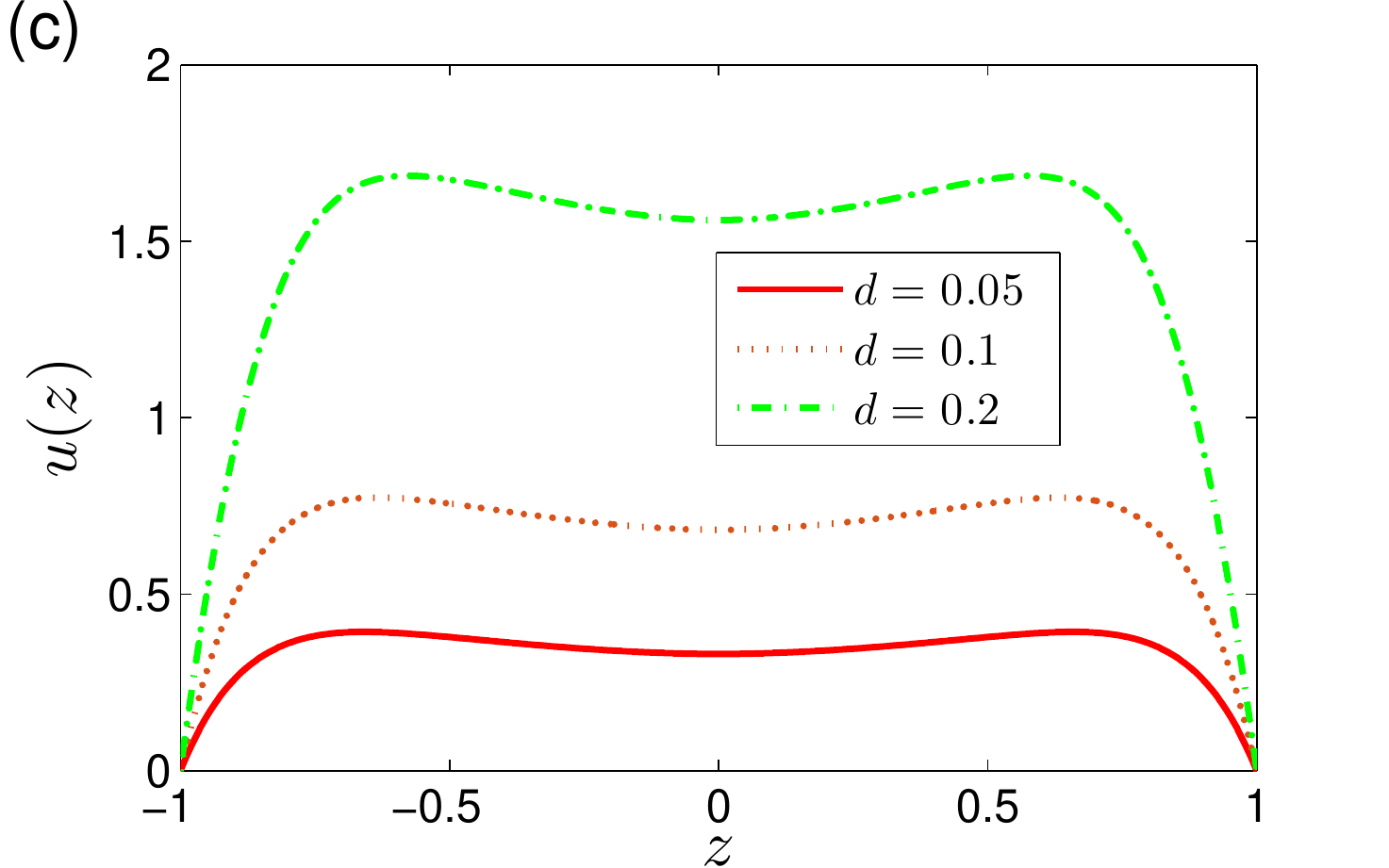}
                \label{fig:k8c}
        \end{subfigure}\hfill
        \begin{subfigure}[t]{0.5\textwidth}
                \includegraphics[width=1\linewidth]{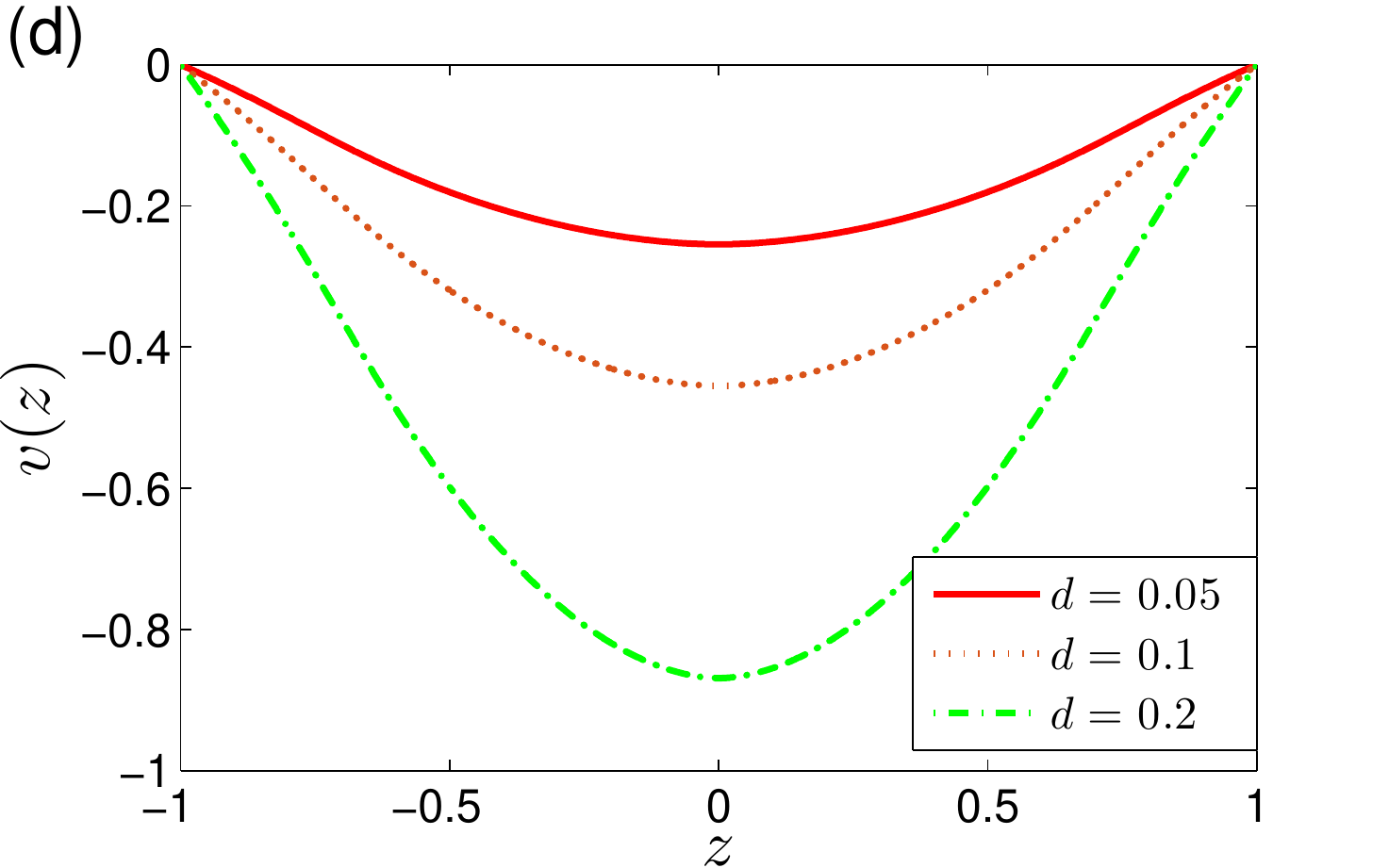}
                \label{fig:k8d}
        \end{subfigure}
 \caption{The variation of (a) $u$-velocity profile for $n= 0.8$, (b) $v$-velocity profile for $n= 0.8$, (c) $u$-velocity profile for $n= 1.2$ and (d) $v$-velocity profile  for $n= 1.2$ for different dimensionless PEL thickness $d$ with $Re_\Omega=5$, $\beta = 0.9$, $\kappa=10$ and $\alpha=2.5$}
\label{fig:k8}
\end{center}
\end{figure}

One of the most important parameters to understand the effect of PEL on the rotational EO flow of power-law fluids is the PEL thickness. The dimensionless PEL thickness is given by $d$ and as $d$ increases the PEL thickness increases. It can be seen from figure \ref{fig:k8} that increasing the grafted layer thickness increases the magnitude of velocity in both $x$ and $y$ directions for both shear thinning and shear thickening fluids. The number of positive ions in the PEL is quite high for larger values of $d$ leading to an increase in electrostatic potential. This  increase in the magnitude of charge within the PEL also increases the electrostatic potential in the EL as the grafting layer dimension is increased which was also observed by Chanda et al. \cite{chandasinhadas}. The increase in the electrostatic potential increases the velocity so much so that despite the extra flow drag in the PEL, the velocity tends to increase with increase in $d$. Therefore, it can be stated that increasing the thickness of the grafted layer can be used to increase both the primary flow velocity magnitude $u$ and secondary flow velocity magnitude $v$ for both shear thinning and shear thickening fluids as seen from figures \ref{fig:k8} (a), (b), (c) and (d). 

\subsection{Effect of rotation of flow rate}

\begin{figure}
\begin{center}
        \begin{subfigure}[t]{0.5\textwidth}
                \includegraphics[width=1\linewidth]{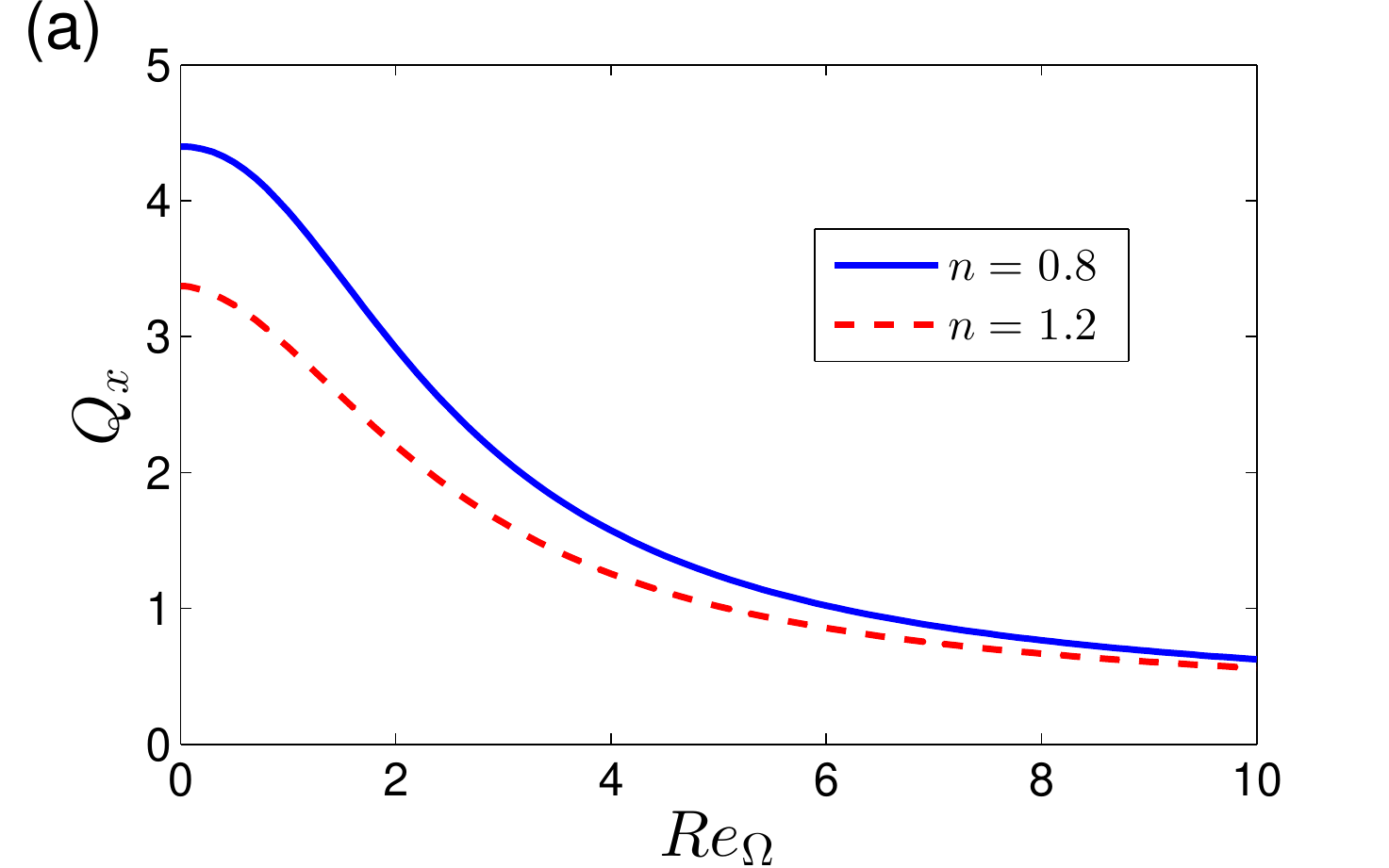}
                \label{fig:k9a}
        \end{subfigure}\hfill
        \begin{subfigure}[t]{0.5\textwidth}
                \includegraphics[width=1\linewidth]{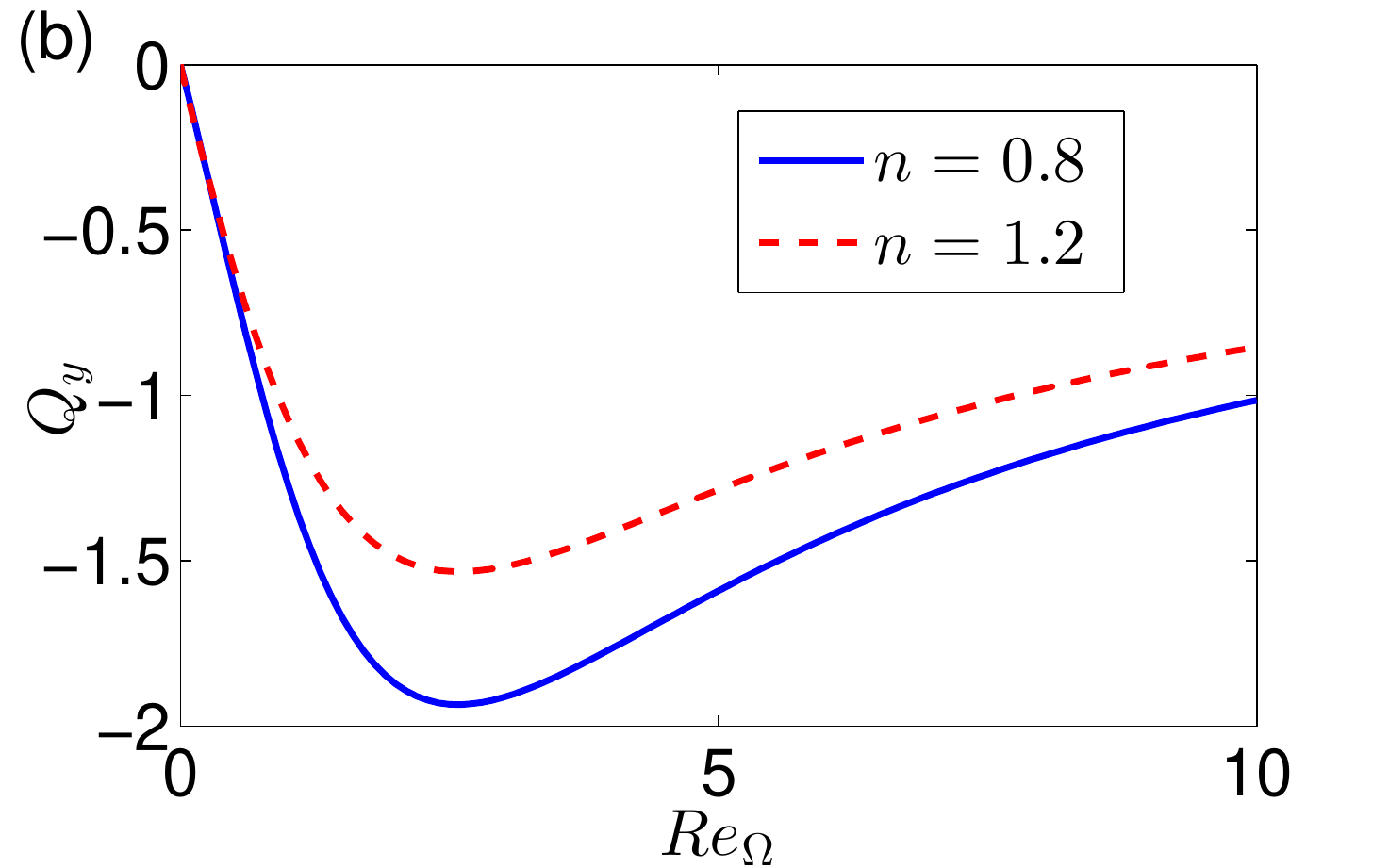}
                \label{fig:k9b}
        \end{subfigure}\\
        \begin{subfigure}[t]{0.5\textwidth}

               \includegraphics[width=1\linewidth]{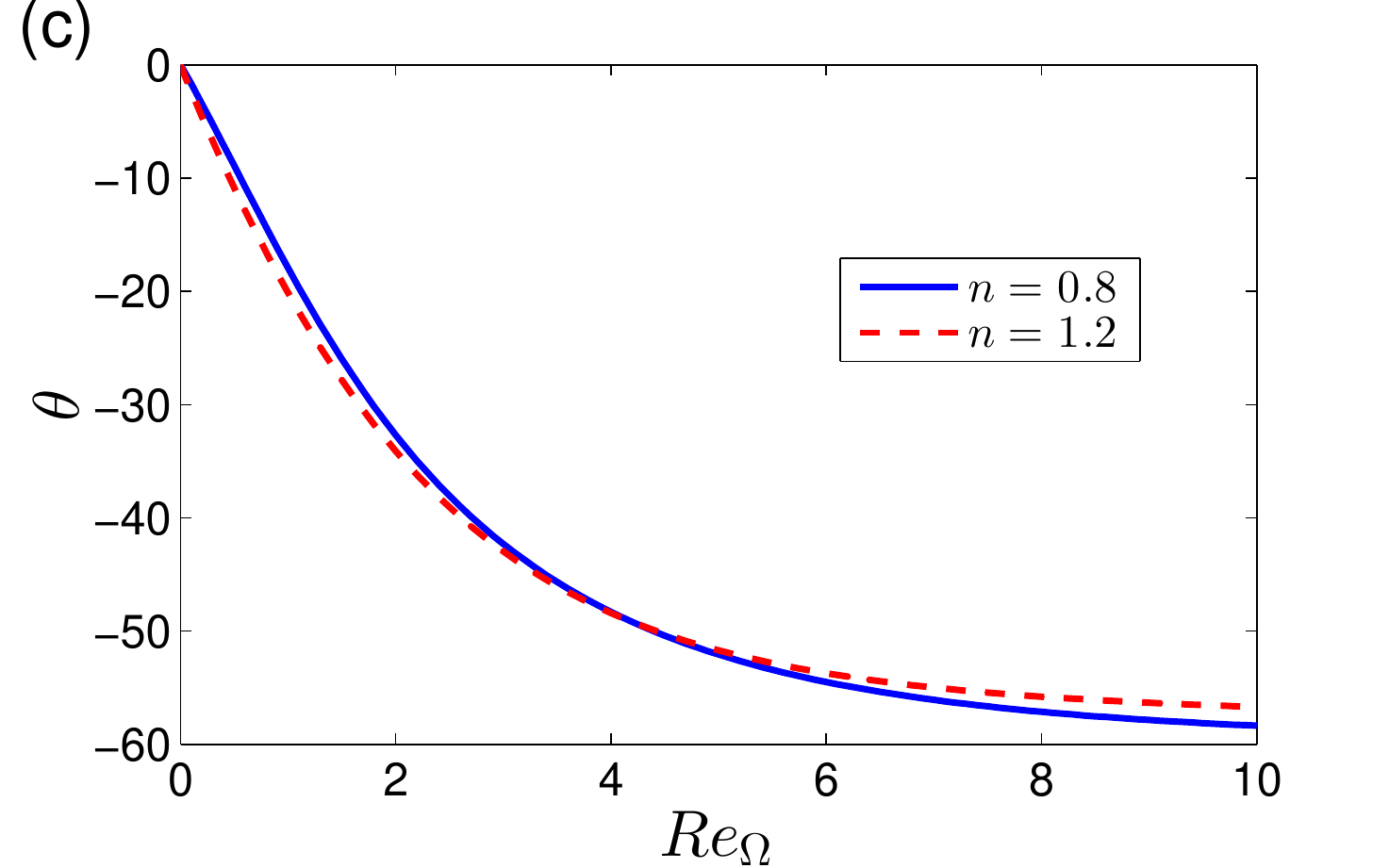}

                \label{fig:k9c}
        \end{subfigure}\hfill
 \caption{The variation of (a) flowrate in the $x$-direction $Q_x$, (b) flowrate in the $y$-direction $Q_y$ and (c) flow angle $\theta$ with $Re_\Omega$ for different power-law index $n$ with $\beta = 0.9$, $\kappa=10$ and $\alpha=2.5$}
\label{fig:k9}
\end{center}
\end{figure}

It is always important to study the flow rate of the fluid as it is simplest measurable quantity in any flow. Here, since there is flow in both $x$ and $y$ directions, we find the flow rate in both the directions as shown in figure 
\ref{fig:k9}. In order to evaluate the flow rates in $x$ and $y$ directions, we use $Q_x = 2 \int_{0}^{1} u dz$ and $Q_y = 2 \int_{0}^{1} v dz$. To find the angle of flow, we use $\theta = \tan^{-1} \left( \dfrac{Q_y}{Q_x} \right)$. We depict in figures \ref{fig:k9} (a), (b) and (c) the flow rates $Q_x$, $Q_y$ and the angle $\theta$ respectively. It can be seen from figure \ref{fig:k9} (a) that the flow rate in the $x$-direction decreases with increase in $Re_\Omega$ for both shear thickening fluid and shear thinning fluid. The rate of decrease of the $x$-direction flow rate reduces with increase in $Re_\Omega$ as the graph flattens out. In figure \ref{fig:k9} (b), we show the $y$ direction flow rate $Q_y$ and we see that $Q_y$ increases in magnitude upto a certain value of $Re_\Omega$ and then starts reducing in magnitude. This is because of interplay of the energy transfer between the $x$ and $y$ momentum equations due to Coriolis effect. In general we see that the magnitude of flow rates is lower of shear thickening fluids as compared to shear thinning fluids. We can therefore choose an optimal speed of rotation where the transverse flow rate is the largest. This can aid in mixing of the fluid in applications where higher mixing is needed during flows through narrow confinements. The variation of the flow angle $\theta$ with $Re_\Omega$ is depicted in \ref{fig:k9} (c). We observe that the flow angle keeps changing with $Re_\Omega$ upto a certain value of $Re_\Omega$ and then becomes almost constant i.e., the flow angle becomes almost independent of $Re_\Omega$ for larger values of $Re_\Omega$. In general, the flow angles for shear thinning fluids is higher than that of shear thickening fluids.

\section{Conclusion}
We have investigated the combined effect of rotational and EO forcing of a power-law fluid throught a soft (PE grafted) narrow channel. We developed a numerical code to solve the transport equations governing the fluid flow. We study the effect of soft PEL grafting on the Coriolis forcing of EOF. The complex interlinked dynamics between the fluid rheology, Coriolis effect and softness effect of the channel has been brought out. We find that each of effect of PE grafting has three separate effects which include the flow drag, the grafting thickness and EDL thickness within the PEL. These effect of fluid rheology and Coriolis forcing on these 3 effects are discussed in detail. We also discussed that the flow rate is significantly affected by the Coriolis forces as well as the fluid rheology and we found that flow rates are typically higher for shear thinning fluids. We believe that studying effect of fluid rheology becomes very important for designing soft channel based Lab-on-CD systems driven by EO forcing and dealing with bio-fluids such as blood, saliva or mucus.

\bibliography{mybibfile}

\end{document}